\documentclass[sigconf]{acmart}
\AtBeginDocument{%
  }

\setcopyright{none}
\renewcommand\footnotetextcopyrightpermission[1]{}

\acmConference[ICSE 2026]{48th International Conference on Software Engineering}{April 2024}{Rio De Janeiro, Brazil}
\usepackage{makecell}
\usepackage{booktabs}
\usepackage{multirow}
\usepackage{amsmath}
\usepackage[table]{xcolor}
\usepackage{listings}
\usepackage{xcolor}
\usepackage{tcolorbox}
\usepackage{xspace}
\usepackage[inline]{enumitem}
\usepackage{verbatim}
\usepackage{soul}
\usepackage[normalem]{ulem}
\renewcommand{\sout}[1]{}
\renewcommand{\st}[1]{}

\lstdefinelanguage{json}{
    basicstyle=\ttfamily\footnotesize,
    numbers=none,
    showstringspaces=false,
    breaklines=true,
    frame=single,
    backgroundcolor=\color{gray!5},
    literate=
     *{0}{{{\color{numb}0}}}{1}
      {1}{{{\color{numb}1}}}{1}
      {2}{{{\color{numb}2}}}{1}
      {3}{{{\color{numb}3}}}{1}
      {4}{{{\color{numb}4}}}{1}
      {5}{{{\color{numb}5}}}{1}
      {6}{{{\color{numb}6}}}{1}
      {7}{{{\color{numb}7}}}{1}
      {8}{{{\color{numb}8}}}{1}
      {9}{{{\color{numb}9}}}{1}
      {:}{{{\color{punct}{:}}}}{1}
      {,}{{{\color{punct}{,}}}}{1}
      {\{}{{{\color{delim}{\{}}}}{1}
      {\}}{{{\color{delim}{\}}}}}{1}
      {[}{{{\color{delim}{[}}}}{1}
      {]}{{{\color{delim}{]}}}}{1},
    morestring=[b]",
    stringstyle=\color{black},
}

\definecolor{numb}{rgb}{0.7,0.0,0.0}
\definecolor{punct}{rgb}{0.3,0.3,0.3}
\definecolor{delim}{rgb}{0.0,0.2,0.4}

\newcommand{\tool}{\textbf{LLM4Perf}\xspace}

\newcommand{\phead}[1]{\vspace{1mm} \noindent {\bf \textit{#1}}}

\newcommand{\rv}[1]{{\leavevmode\color{red}#1}}
\renewcommand{\rv}[1]{#1}

\newcommand{\greybox}[1]{
\vspace{0.05cm}
    \begin{tcolorbox}[
        left=2pt, right=2pt, top=2pt, bottom=2pt,
        boxrule=0.2mm,
        leftrule=1mm,
        arc=0mm,
        colframe=black!80!white, % 边框颜色
        colback=black!2!white, % 文本框内颜色
        colbacktitle=black!50!white % 标题底色
    ]
    {#1}
    \end{tcolorbox}
\vspace{0.05cm}
}

\setlist[itemize]{leftmargin=2em}

\begin{document}

%%
%% The "title" command has an optional parameter,
%% allowing the author to define a "short title" to be used in page headers.
\title{LLM4Perf: Large Language Models Are Effective Samplers for Multi-Objective Performance Modeling}

%%
%% The "author" command and its associated commands are used to define
%% the authors and their affiliations.
%% Of note is the shared affiliation of the first two authors, and the
%% "authornote" and "authornotemark" commands
%% used to denote shared contribution to the research.
% \author{Ben Trovato}
% \authornote{Both authors contributed equally to this research.}
% \email{trovato@corporation.com}
% \orcid{1234-5678-9012}
% \author{G.K.M. Tobin}
% \authornotemark[1]
% \email{webmaster@marysville-ohio.com}
% \affiliation{%
%   \institution{Institute for Clarity in Documentation}
%   \city{Dublin}
%   \state{Ohio}
%   \country{USA}
% }

\author{Xin Wang}
\email{xwang496@connect.hkust-gz.edu.cn}
\affiliation{%
  \institution{The Hong Kong University of Science and Technology (Guangzhou)}
  \city{Guangzhou}
  \country{China}
}

\author{Zhenhao Li}
\email{lzhenhao@yorku.ca}
\affiliation{%
  \institution{York University}
  \city{Toronto}
  \country{Canada}
}

\author{Zishuo Ding\textsuperscript{*}}
\email{zishuoding@hkust-gz.edu.cn}
\affiliation{%
  \institution{The Hong Kong University of Science and Technology (Guangzhou)}
  \city{Guangzhou}
  \country{China}
}

\thanks{\textsuperscript{*}Corresponding authors.}

% \author{Lars Th{\o}rv{\"a}ld}
% \affiliation{%
%   \institution{The Th{\o}rv{\"a}ld Group}
%   \city{Hekla}
%   \country{Iceland}}
% \email{larst@affiliation.org}

% \author{Valerie B\'eranger}
% \affiliation{%
%   \institution{Inria Paris-Rocquencourt}
%   \city{Rocquencourt}
%   \country{France}
% }

% \author{Aparna Patel}
% \affiliation{%
%  \institution{Rajiv Gandhi University}
%  \city{Doimukh}
%  \state{Arunachal Pradesh}
%  \country{India}}

% \author{Huifen Chan}
% \affiliation{%
%   \institution{Tsinghua University}
%   \city{Haidian Qu}
%   \state{Beijing Shi}
%   \country{China}}

% \author{Charles Palmer}
% \affiliation{%
%   \institution{Palmer Research Laboratories}
%   \city{San Antonio}
%   \state{Texas}
%   \country{USA}}
% \email{cpalmer@prl.com}

% \author{John Smith}
% \affiliation{%
%   \institution{The Th{\o}rv{\"a}ld Group}
%   \city{Hekla}
%   \country{Iceland}}
% \email{jsmith@affiliation.org}

% \author{Julius P. Kumquat}
% \affiliation{%
%   \institution{The Kumquat Consortium}
%   \city{New York}
%   \country{USA}}
% \email{jpkumquat@consortium.net}

%%
%% By default, the full list of authors will be used in the page
%% headers. Often, this list is too long, and will overlap
%% other information printed in the page headers. This command allows
%% the author to define a more concise list
%% of authors' names for this purpose.
\renewcommand{\shortauthors}{Trovato et al.}

%%
%% The abstract is a short summary of the work to be presented in the
%% article.
\begin{abstract}
The performance of modern software systems is critically dependent on their complex configuration options. Building accurate performance models to navigate this vast space requires effective sampling strategies, yet existing methods often struggle with multi-objective optimization and cannot leverage semantic information from documentation. The recent success of Large Language Models (LLMs) motivates the central question of this work: \textbf{Can LLMs serve as effective samplers for multi-objective performance modeling?} To explore this, we present a comprehensive empirical study investigating the capabilities and characteristics of LLM-driven sampling. We design and implement \tool, a feedback-based framework, and use it to systematically evaluate the LLM-guided sampling process across four highly-configurable, real-world systems. Our study reveals that the LLM-guided approach outperforms traditional baselines in most cases. Quantitatively, \tool achieves the best performance in nearly \rv{68.8\% (77 out of 112)} \sout{72\% (81 out of 112)} of all evaluation scenarios, demonstrating its superior effectiveness. We find this effectiveness stems from the LLM's dual capabilities of configuration space pruning and feedback-driven strategy refinement. The effectiveness of this pruning is further validated by the fact that it also improves the performance of the baseline methods in nearly \rv{91.5\% (410 out of 448)} \sout{87\% (246 out of 336)} of cases. Furthermore, we show how the LLM choices for each component and hyperparameters within \tool affect its effectiveness. Overall, this paper provides strong evidence for the effectiveness of LLMs in performance engineering and offers concrete insights into the mechanisms that drive their success.
\end{abstract}

\maketitle

\section{Introduction}
\label{sec:introduction}
Modern software systems are highly configurable, offering users flexible control through a wide range of options, such as algorithmic choices,  thresholds, and resource settings. While this configurability enhances adaptability, it also presents challenges to identifying configurations that ensure high performance. The combination of options leads to a combinatorial explosion in the configuration search space, where each configuration (i.e., a specific combination of option values)  can have an unpredictable effect on key performance metrics like latency and memory usage, directly impacting the user experience. As a result, developers rely on performance testing to find effective configurations. However, exhaustively evaluating all combinations is computationally infeasible. This challenge is further compounded by complex trade-offs, where improving one objective may degrade another, making it difficult to deliver consistently high-performing systems.

A common approach to managing configuration complexity is to build predictive models that estimate system performance based on a small sample of tested configurations. Research in this area has advanced along two directions.
On one hand, efforts have focused on the predictive models themselves, where researchers have proposed a range of approaches, including traditional machine learning techniques, as well as deep learning models like DeepPerf~\cite{haDeepPerfPerformancePrediction2019} and Perf-AL~\cite{shuPerfALPerformancePrediction2020}. On the other hand, researchers have recognized that the accuracy of these models depends heavily on the quality of the sampled configuration data. This has led to the development of various sampling strategies to guide configuration selection. For example, diversity-driven methods aim to maximize coverage of the configuration space~\cite{kaltenecker_DistanceBasedSamplingSoftware_2019}, search-based methods seek high-performing configurations~\cite{guo_GeneticAlgorithmOptimized_2011, lemieux_PerfFuzzAutomaticallyGenerating_2018, xiangSearchbasedDiverseSampling2022}, and feedback-based approaches leverage feedback loops to iteratively refine the sampling process~\cite{siegmund_ScalablePredictionNonfunctional_2011, henardCombiningMultiObjectiveSearch2015, xiaCoMSAModelingDrivenSampling2023}.

However, existing sampling methods have several limitations. First, they typically treat all configuration options as equally important, failing to leverage domain knowledge, such as developer documentation, to identify and prune performance-insensitive options. These options expand the configuration space used for sampling, thereby hindering the sampling effectiveness. Second, diversity-based strategies, though effective for ensuring broad coverage of the configuration space~\cite{kaltenecker_DistanceBasedSamplingSoftware_2019, xiangSearchbasedDiverseSampling2022}, may be less suitable for performance modeling tasks. In practice, different configurations may still exhibit similar performance, leading to redundant measurements. While feedback-driven strategies~\cite{nairFindingFasterConfigurations2020, guo_GeneticAlgorithmOptimized_2011, xiaCoMSAModelingDrivenSampling2023} address this by adapting the sampling process based on observed performance results, most are designed to optimize a single performance metric. This poses a challenge in real-world performance tuning, which often involves balancing complex multi-objective trade-offs, such as latency versus memory usage. Applying these single-objective methods in such cases requires running the entire sampling process separately for each metric, which is inefficient and fails to capture the interactions between different performance goals.

Recent advances in LLMs have motivated extensive research into applying LLMs across a wide range of software engineering tasks, such as code generation~\cite{wang2025codevisionary, yang2025code, chen2024code, chen2025reasoning}, detecting software defects~\cite{hassan2024state, wang2025defects4log}, program repair~\cite{wang2025defects4c, yin2024thinkrepair}, as well as those related to performance analysis and optimization~\cite{wangIdentifyingPerformanceSensitiveConfigurations2024, pengPerfCodeGenImprovingPerformance2024, pengSysLLMaticLargeLanguage2025, cuiLargeLanguageModels2025, peng2025coffe}. Inspired by this trend,  we would like to \textbf{explore whether the reasoning and generalization capabilities of LLMs can be leveraged to overcome the aforementioned limitations by incorporating domain knowledge, utilizing feedback, and supporting multi-objective optimization.} To this end,
we conduct a comprehensive study to assess the feasibility and effectiveness of using LLMs as configuration samplers for performance modeling.

In this study, we design and implement \tool, a novel framework for LLM-driven configuration sampling. In \tool, the LLM analyzes performance documentation as domain knowledge to prune the configuration space and uses iterative feedback to refine the sampling strategy, thereby guiding the sampling toward more effective regions of the configuration space. The framework is specifically designed to handle multi-objective performance scenarios and aims to produce a sampling strategy that is both effective and efficient. Meanwhile, to support our evaluation, we construct a new multi-objective performance dataset based on four real-world systems. Unlike existing datasets, ours records performance metrics under both the full configuration space and a pruned space where performance-insensitive options are removed. This design allows us to examine the effects of configuration space pruning and assess the framework’s ability to handle multi-objective optimization.

Our empirical evaluation, conducted using the \tool framework and our newly constructed dataset, addresses four key research questions:
\begin{enumerate*}[label=(\roman*)]
    \item \textbf{Effectiveness (RQ1):} We compare \tool against existing sampling baselines. The results show that both traditional machine learning model (i.e., XGBoost) and deep learning models (i.e., DeepPerf), when trained on configuration samples generated by \tool, generally achieve better performance across multiple metrics.
    \item \textbf{Core Mechanisms (RQ2):} We analyze the underlying factors contributing to \tool’s effectiveness and identify two key sources: the LLM’s ability to prune the configuration space and its capability to iteratively refine the sampling strategy based on feedback.
    \item \textbf{Impact of LLM Selection (RQ3):} We evaluate the impact of each component’s LLM choice on the overall effectiveness of \tool. The results indicate that equipping reasoning‑intensive components with high‑capability LLMs improves overall effectiveness, whereas components with modest reasoning requirements can employ smaller models without perceptible degradation. 
    \item \textbf{Impact of Hyperparameter (RQ4):} We investigate how \tool’s hyperparameters settings influence its behavior.
\end{enumerate*}

In summary, this paper makes the following contributions:
\begin{itemize}
    \item We propose \tool, an LLM-driven framework that leverages LLMs to prune the configuration space and adaptively guide the sampling process using iterative feedback. \tool is specifically designed to support multi-objective performance modeling. Our replication package is publicly available~\cite{replication_package}.
    \item We construct a new multi-objective performance dataset covering four real-world systems and use it to evaluate \tool across four dimensions: effectiveness, core mechanisms, the influence of LLM selection, and the impact of hyperparameter settings.
    \item Our results show that \tool enables more accurate and stable performance modeling compared to existing baselines.
    % generalizes across multiple open-source LLMs, and exhibits practical performance–cost trade-offs, making it suitable for real-world adoption.
    % \textbf{A publicly available \todo{replication package}~\cite{replication_package},} including newly created multi-metric performance datasets for four real-world systems and source code of \tool.
\end{itemize}

\textbf{Paper Organization.} The rest of this paper is organized as follows. Section~\ref{sec:related_work} surveys prior work on configuration sampling, performance modeling, and LLM‑based performance tasks. Section~\ref{sec:methodology} presents the design and implementation of \tool. Section~\ref{sec:experiment_setup} outlines the experimental setup, and Section~\ref{sec:research_question} states and discusses our four research questions. Section~\ref{sec:threat} analyzes threats to validity. Finally, Section~\ref{sec:conclusion} concludes the paper.

\section{Related Work}
\label{sec:related_work}
\subsection{Configuration Sampling Strategies}
% \zhl{Use simple past tense in this section }
Accurately modeling the performance of highly configurable systems requires effective sampling strategies. The sampling strategies have evolved from simple random approaches to more sophisticated, model-guided techniques.

The most prevalent approach is random sampling, which the literature distinguishes into two main types: simple random sampling~\cite{Jeho_FindingNearOptimal_2017, haPerformanceInfluenceModelHighly2019, javidian_TransferLearningPerformance_2019, krishna_WhenceLearnTransferring_2020} and feature-size heuristic sampling~\cite{guo_VariabilityawarePerformancePrediction_2013, guo_DataefficientPerformanceLearning_2018, haDeepPerfPerformancePrediction2019, nairFindingFasterConfigurations2020, shuPerfALPerformancePrediction2020}. While straightforward to implement, random approaches are often inefficient as they may fail to capture performance-critical configurations. To improve coverage, diversity-driven sampling~\cite{siegmund_PredictingPerformanceAutomated_2012, lillacka_ImprovedPredictionOf_2013, bao_AutoConfigAutomaticConfiguration_2018, kaltenecker_DistanceBasedSamplingSoftware_2019} uses distance metrics (e.g., Manhattan distance~\cite{krause_TaxicabGeometryAdventure_2012}) to select configurations that are maximally dissimilar, though this approach didn't explicitly consider performance impact.

In contrast, search-based strategies like Genetic Algorithms (GAs)~\cite{guo_GeneticAlgorithmOptimized_2011, luoMiningPerformanceRegression2016, lemieux_PerfFuzzAutomaticallyGenerating_2018} treated sampling as an optimization problem. They use an evolutionary search guided by a performance-based fitness function to find high-performing configurations, but risked converging to local optima. Feedback-based sampling used a performance model in a feedback loop to guide the selection. This included solver-based sampling~\cite{siegmund_ScalablePredictionNonfunctional_2011, henardCombiningMultiObjectiveSearch2015, xiangSearchbasedDiverseSampling2022}, which used constraint solvers to find valid configurations, as well as methods that either exploited the model's predictions~\cite{nairFindingFasterConfigurations2020} or explored areas of high model uncertainty~\cite{xiaCoMSAModelingDrivenSampling2023}.

While these methods have significantly advanced the field, they often focused on a single performance metric and didn't leverage the rich semantic information available in developer documentation. Our work aims to bridge this gap by using LLMs to create a more holistic and context-aware sampling framework.
% \subsection{Performance Modeling}

\subsection{Performance Modeling Methods}

The task of building predictive models for performance based on configuration data was approached with a diverse set of techniques. Initial efforts in this domain centered on classical machine learning algorithms, which included foundational methods like linear regression~\cite{siegmundPerformanceinfluenceModelsHighly2015, kangyong-bin_UnderstandingImprovingOntology_2020, siegmundnorbert_SPLConqueror_2012, sun_AutomatedPerformanceModeling_2020}, as well as techniques such as tree-based models~\cite{hsu_ArrowLowLevelAugmented_2018, nairFindingFasterConfigurations2020, chen_XGBoostScalableTree_2016} and Fourier-based learning~\cite{haPerformanceInfluenceModelHighly2019, zhang_PerformancePredictionConfigurable_2015}. 

A drawback of these classical approaches, however, is their limited effectiveness on the small and sparse datasets typically generated from expensive performance measurements~\cite{haDeepPerfPerformancePrediction2019}. This challenge, particularly the issue of feature sparsity, prompted a shift towards deep learning solutions, with many recent studies applying deep neural networks to the task of performance prediction~\cite{falch_MachineLearningBased_2015, haDeepPerfPerformancePrediction2019, moradi_PerformancePredictionDynamic_2019, kim_P4PhasebasedPower_2017, marathe_PerformanceModelingResource_2017, nemirovsky_MachineLearningApproach_2017, shuPerfALPerformancePrediction2020}. Notable examples in this area included HINNPerf~\cite{chengHINNPerfHierarchicalInteraction2023}, which employed embeddings with hierarchical regularization, and \texttt{Perf-AL}~\cite{shuPerfALPerformancePrediction2020}, which used an adversarial learning architecture. A particularly influential model, \texttt{DeepPerf}~\cite{chengHINNPerfHierarchicalInteraction2023}, became a state-of-the-art method by using a deep neural network with L1 regularization to specifically address feature sparsity.

\begin{figure*}[tbp]
    \centering
    \includegraphics[width=0.9\textwidth]{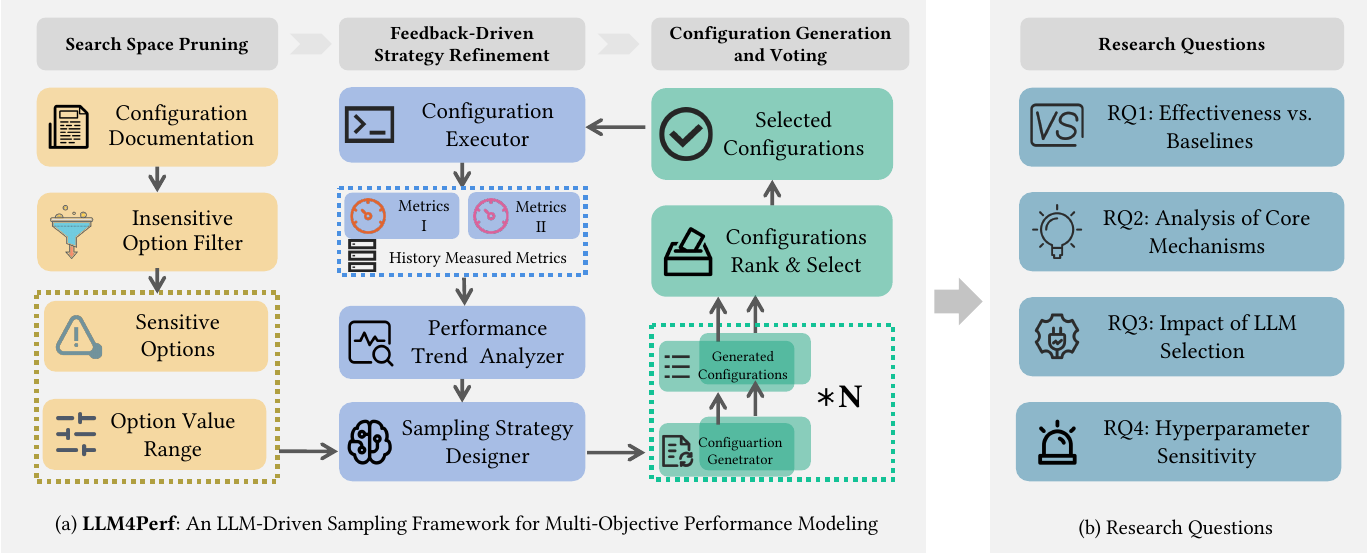}
    \caption{An overview of our study.}
    \label{fig:overview}
\end{figure*}
\subsection{LLM-based Performance Tasks}
In recent years, LLMs have emerged as a powerful tool in many software engineering tasks. Their remarkable ability to understand complex, unstructured data, such as natural language documentation and source code, makes them uniquely suited for challenges in software performance engineering that were previously difficult to automate.

Consequently, researchers began to explore the potential of LLMs for performance-related tasks. One line of research focused on automated code optimization. For instance, ~\citet{pengSysLLMaticLargeLanguage2025} proposed SysLLMatic, a system that integrated an LLM with profiling-guided feedback to automatically rewrite software code. Similarly, ~\citet{pengPerfCodeGenImprovingPerformance2024} introduced PerfCodeGen, a framework that enhanced code performance through self-refinement, incorporating runtime feedback from test executions to iteratively improve the generated code. The application of LLMs for code optimization was also explored in specialized domains. ~\citet{cuiLargeLanguageModels2025} evaluated several state-of-the-art LLMs on High-Performance Computing (HPC) tasks, finding that while LLMs were proficient at basic transformations, they still struggled with the complex control and data flows inherent in HPC code. Another emerging direction, which is more closely related to our work, applied LLMs to the configuration space. ~\citet{wangIdentifyingPerformanceSensitiveConfigurations2024} introduced PerfSense, a framework that used LLM agents to identify performance-sensitive configurations. PerfSense simulated interactions between developers and performance engineers, employing prompting techniques like prompt chaining and Retrieval-Augmented Generation (RAG) to analyze configuration descriptions and pinpoint influential parameters. 
While PerfSense focuses on identifying sensitive configurations, we use the LLM to dynamically guide the entire iterative sampling process based on performance feedback.
Peng et al.~\cite{peng2025coffe} proposed a benchmark for evaluating the efficiency of LLM-generated code.

While these works demonstrated the potential of LLMs in performance engineering, they primarily focused on either direct code optimization or identifying sensitive parameters in a pre-analysis step. The challenge of leveraging an LLM to guide the iterative sampling process for building accurate, multi-metric performance models remains largely unexplored. Our work addresses this specific gap by proposing a framework where an LLM actively participates in a feedback loop to dynamically refine the sampling strategy over time.

% \section{Preliminary Study}
% \input{content/04pre study}
\section{Methodology}
\label{sec:methodology}

\st{As shown in Figure 1(a), LLM4Perf consists of three phases. First, in the Search Space Pruning phase, it leverages domain knowledge extracted from project documentation to filter out performance-insensitive options, thereby constructing a focused configuration space composed of only sensitive options. Second, during Feedback-Driven Strategy Refinement, LLM4Perf iteratively executes sampled configurations, records performance metrics, and employs an LLM-based analyzer to uncover trends and trade-offs across multiple objectives. These insights are used to update the sampling strategy. Finally, in the Diverse Configuration Generation and Voting phase, multiple LLMs generate candidate configurations based on the refined strategy. A voting module deduplicates, ranks, and selects the next batch for execution. This process repeats until the sampling budget is exhausted, yielding a configuration set for training performance models.}

\rv{As shown in Figure\ref{fig:overview}(a), \tool consists of five core components: \begin{itemize} \item The \textbf{Non-Impactful Option Filter}, which analyzes documentation to prune the initial search space by removing performance-insensitive options. \item The \textbf{Performance Trend Analyzer}, an LLM agent that uncovers performance trends, anomalies, and trade-offs from historical execution data. \item The \textbf{Sampling Strategy Designer}, which uses the analyzer's insights to create a high-level plan for the next sampling iteration. \item The \textbf{Configuration Generators}, a set of parallel LLMs that generate diverse candidate configurations adhering to the given strategy. \item The \textbf{Configuration Voting} module, which aggregates, deduplicates, and ranks the candidates to select the final batch for execution. \end{itemize}}

\rv{These components operate in a three-phase process. The workflow begins with the \textbf{Search Space Pruning} phase, executed by the \textbf{Non-Impactful Option Filter}. This is followed by an iterative feedback loop. In the \textbf{Feedback-Driven Strategy Refinement} phase, the \textbf{Performance Trend Analyzer} and \textbf{Sampling Strategy Designer} collaborate to update the sampling plan based on observed metrics. Finally, in the \textbf{Diverse Configuration Generation and Voting} phase, the \textbf{Configuration Generators} and \textbf{Configuration Voting} module produce the next batch of configurations. This process repeats until the sampling budget, \textbf{defined as the total number of configurations to be executed (e.g., a budget of 70 samples)}, is exhausted, yielding a configuration set for training performance models.}

\subsection{Search Space Pruning}

\rv{The first phase addresses the high-dimensionality challenge: many configuration options exert little or no influence on performance. The domain knowledge that distinguishes performance-sensitive from insensitive options resides in developer documentation, which traditional samplers overlook. Our methodology therefore makes a key assumption: \textbf{that documentation for the system's configuration options is available}.

To leverage this documentation, we first process the \textbf{Configuration Documentation} (which may be in formats like man pages, HTML, or Markdown) into a structured \texttt{JSON} format(Listing~\ref{lst:lrzip_json_example}). This preprocessing step standardizes the input, extracting the name, description, and value constraints for each parameter. For example, based on the documentation for the lrzip compression utility, the ``algorithm'' option would be extracted alongside a non-performance-critical option like ``-N'':
}

\begin{lstlisting}[language=json, caption={\rv{Example JSON representation of configuration documentation from the \texttt{lrzip} utility.}}, label={lst:lrzip_json_example}]
[
  {
    "name": "algorithm",
    "description": "Specifies the compression algorithm. Available options include: -g (Gzip, balanced speed), -l (LZO, ultra fast), -z (ZPAQ, highest ratio but extremely slow)."
  },
  {
    "name": "-N",
    "description": "Sets process priority (nice value from -20 to 19). This parameter does not affect compression speed or ratio."
  }
]
\end{lstlisting}

\rv{This structured data is then processed by the \textbf{Non-Impactful Option Filter}, an LLM agent prompted to act as a performance expert. The agent semantically analyzes each parameter's description. Based on this analysis, it prunes options unlikely to have a significant performance impact. In the example above, the agent would discard \texttt{-N}, as its description explicitly states it ``does NOT speed up or slow down compression.'' Conversely, it would identify \texttt{algorithm} as performance-sensitive because its description mentions trade-offs in speed and compression ratio (e.g., ``ultra fast'' vs. ``extremely slow'').

The output is a curated list of \textbf{Sensitive Options}. This list, combined with the extracted value ranges, forms a pruned search space that directs all subsequent sampling efforts toward the most performance-critical regions.}

\subsection{Feedback-Driven Strategy Refinement}

While the initial pruning phase effectively reduces noise using static documentation, it may not fully capture complex runtime interactions among configuration options. To address this limitation, the second phase of our framework introduces a dynamic feedback loop that progressively refines the sampling strategy based on empirical performance data collected during execution.

In the first iteration, \textbf{\tool} generates an initial batch of configurations from the pruned search space without relying on any prior measurements. The \textbf{Configuration Executor} then evaluates these configurations and stores the resulting multi-objective performance metrics in the \textbf{History Measured Metrics} database. From the second iteration onward, this historical data serves as feedback for adaptive refinement. The \textbf{Performance Trend Analyzer} examines the evolving dataset to identify correlations and trade-offs among configuration parameters and performance outcomes. For instance, it may observe that specific combinations, such as using the \texttt{-z} algorithm with smaller window sizes, increase compression ratio but also lead to higher memory usage.

Based on the insights obtained from this analysis, the \textbf{Sampling Strategy Designer} formulates a refined sampling plan expressed in natural language. This updated strategy provides explicit guidance for the next round of configuration generation, emphasizing underexplored but potentially high-impact regions of the configuration space.

Through this iterative feedback process, \textbf{\tool} incrementally enhances its understanding of the configuration–performance relationship and directs subsequent sampling toward performance-critical areas. This design enables the framework to continuously improve both the efficiency and the predictive accuracy of performance modeling across iterations.

\subsection{Configuration Generation and Voting}

The final phase operationalizes the high-level strategy generated in the previous step into a concrete set of candidate configurations. To maintain diversity and mitigate the inherent randomness of LLMs, \textbf{\tool} employs a two-step process consisting of \textbf{parallel generation} and \textbf{voting-based selection}.

First, the updated strategy is broadcast to multiple \textbf{Configuration Generators} running in parallel. Each generator produces a set of candidate configurations aligned with the strategy’s constraints. This parallel design ensures broad exploration of the configuration space while maintaining alignment with the intended sampling focus.

Second, the \textbf{Configuration Voting} module aggregates all candidate lists, removes duplicates, and ranks the configurations according to their frequency of occurrence across different generators. The top-ranked configurations are selected as the \textbf{Final Candidate Set} for execution in the next feedback cycle. This ensemble-style generation and voting process reduces sampling instability, enhances the representativeness of the selected configurations, and yields more reliable performance models.

\subsection{Running Example}
\rv{
We illustrate this three-phase process using a running example of configuring the \texttt{lrzip} compression utility, with a total sampling budget of 20 configurations. The goal is to model performance across three objectives: \textbf{compression time}, \textbf{compression speed}, and \textbf{max memory}.

\noindent \textbf{Phase 1: Pruning.} First, in the Search Space Pruning phase, the \textbf{Non-Impactful Option Filter} analyzes the documentation. It identifies \texttt{algorithm} as performance-sensitive (description mentions ``ultra fast'' vs. ``extremely slow'') but prunes the \texttt{-N} (nice value) option, as its documentation explicitly states it ``does NOT speed up or slow down compression.'' The output is a pruned search space focused on sensitive options like \texttt{algorithm}, \texttt{-w} (window size), \texttt{-p} (processors), and \texttt{-L} (level).

\noindent \textbf{Phases 2 \& 3: Iterative Refinement and Generation (Iteration 1).} The process then enters the iterative loop. For the first iteration, the \textbf{Sampling Strategy Designer}, having no prior data, creates an initial \texttt{exploration strategy} to ``cover all 5 algorithms at extreme parameter values'' (e.g., min/max window size and processor counts). This strategy is sent to parallel \textbf{Configuration Generators}, and the Configuration Voting module selects the first batch of configurations, including \texttt{lrzip -b -L 9 -w 81 -p 4} and \texttt{lrzip -g -L 9 -w 1 -p 1}. These are executed, and their metrics are saved (e.g., \texttt{-b} takes 21.2s; \texttt{-g} takes 40.3s).

\noindent \textbf{Phases 2 \& 3: Iterative Refinement and Generation (Iteration 2).} For the second iteration, the \textbf{Performance Trend Analyzer} examines this new historical data. It generates an analysis (matching the JSON's \texttt{performance analysis}): it reports that \texttt{algorithm -g} is an anomaly, showing ``exceptionally high time (40.27 seconds),'' while \texttt{-b} and \texttt{-n} perform well. It also hypothesizes that \texttt{Compression level (-L)} has ``low sensitivity'' but that \texttt{-p} and \texttt{-w} ``interact strongly.'' Based on this analysis, the \textbf{Sampling Strategy Designer} refines the plan: ``Focus on the untested \texttt{-z} (ZPAQ) algorithm, test intermediate values for \texttt{-p} and \texttt{-w}, and de-prioritize \texttt{-L} by fixing it to 8.'' The \textbf{Generators} and Voting module then produce the next batch based on this new strategy, such as \texttt{lrzip -z -L 8 -w 41 -p 3}.

This feedback loop of analysis, strategy refinement, and generation continues until the 20-sample budget is exhausted, resulting in a diverse and informative set of performance measurements.
}

\section{Experiment Setup}
\label{sec:experiment_setup}
\subsection{Dataset Construction}
Many existing sampling methods for performance modeling are designed to optimize for a single objective. Consequently, the publicly available datasets from these studies typically feature only one performance metric. This presents a significant limitation, as real-world software tuning requires balancing multiple, often conflicting, performance goals. To properly evaluate our framework's ability to handle such multi-metric scenarios, we first construct a new dataset from four highly configurable, open-source projects: \textbf{LRZIP}, \textbf{JavaGC}, \textbf{SQLite}, and \textbf{X264}.

\phead{Dual Configuration Spaces.}
To support our later analysis of performance-insensitive configurations (RQ2, Section~\ref{sec:rq2}), we need to prepare two distinct versions of the configuration space for each subject system. The first version is the \textit{full} space, which includes all original options. To create the second, \textit{pruned} space, we employ LLM to analyze the developer documentation and identify options that were unlikely to impact performance . These identified options are then excluded, resulting in a smaller, more focused configuration space. Preparing both datasets allows us to directly measure the benefits of this LLM-guided space pruning.

% \phead{Performance Metrics.}
% To capture realistic performance trade-offs, we select two distinct types of metrics for each system. The first metric is a general measure of \textbf{temporal performance} reflecting the system's primary task efficiency (e.g., Compression Time, Encoding Time). The second is a \textbf{domain-specific metric} that represents a critical quality or resource constraint of direct concern to end-users, such as memory consumption (e.g., Max Memory) or output quality (e.g., Video Quality (PSNR)).

\phead{Measurement Environment.}
To ensure the reliability of our dataset, all performance measurements are collected in a controlled environment. Each configuration is executed on a virtual machine running Ubuntu 20.04 with 2 vCPUs and 4 GB of RAM. To mitigate measurement noise from system fluctuations, each test run is repeated 10 times, and we record the average of these runs as the final performance value. A summary of the projects, the size of their configuration spaces, and the specific performance metrics we collected is provided in Table~\ref{tab:projects-metrics}.
\begin{table}[h!]
\centering
\caption{Subject Systems and Their Performance Metrics}
\label{tab:projects-metrics}
\resizebox{\columnwidth}{!}{
\begin{tabular}{lllll}
\toprule
\textbf{Project} & \textbf{Metric 1} & \textbf{Metric 2} & \textbf{Full Space Size} & \textbf{Pruned Space Size} \\ 
\midrule
LRZIP       & Compression Time   & Max Memory           & 1,200  & 200    \\
JavaGC      & Collection Time    & Average Pause Time   & 6,240  & 1,560  \\ 
SQLite      & Response Time      & Max Memory           & 9,216  & 1,536  \\
X264        & Encoding Time      & Video Quality(PSNR)  & 13,824 & 3,456  \\
\bottomrule
\end{tabular}
}
\end{table}

\subsection{Baselines}
To provide a comprehensive comparison, we evaluate our method against five baseline strategies that represent distinct families of sampling approaches.

\begin{itemize}
    \item \textbf{Random Sampling~\cite{haPerformanceInfluenceModelHighly2019}:} Selects configurations uniformly at random from the configuration space.
    \item \textbf{Flash~\cite{nairFindingFasterConfigurations2020}:} A sequential model-based method that iteratively selects the configuration predicted to yield the best performance.
    \item \textbf{Genetic Sampling~\cite{guo_GeneticAlgorithmOptimized_2011}:} An evolutionary algorithm that uses crossover and mutation to evolve a population of configurations toward better performance.
    \item \textbf{CoMSA~\cite{xiaCoMSAModelingDrivenSampling2023}:} A model-driven sampler that prioritizes configurations with the highest prediction uncertainty to improve model accuracy.
    \item \textbf{NSBS~\cite{xiangSearchbasedDiverseSampling2022}:} A diversity-driven sampler that uses a distance metric to select configurations most dissimilar from those already sampled.
    \rv{
    \item \textbf{EHVI~\cite{yang_MultiObjectiveBayesianGlobal_2019}:} 
    A Gaussian Process (GP)-based multi-objective method that selects configurations based on their Expected Hypervolume Improvement (EHVI) to expand the Pareto front.
    \item \textbf{TSEMO~\cite{luo_EvolutionaryOptimizationExpensive_2019}:} 
    A Thompson Sampling-based multi-objective algorithm that uses GP posteriors and evolutionary search to probabilistically explore the Pareto front.}
    \item \sout{\textbf{NSGA-II~\cite{verma_ComprehensiveReviewNSGAII_2021}:}} \rv{\textbf{NSGA-III~\cite{deb_EvolutionaryManyObjectiveOptimization_2014}}} A widely-used multi-objective evolutionary algorithm that evolves a population to find the set of optimal trade-offs between conflicting objectives.
\end{itemize}
% \vspace{-1mm}
\subsection{Performance Modeling Methods}

Once a set of configurations is sampled, a performance model is trained on the resulting data to predict the performance of untested configurations. The choice of this model directly influences the final prediction accuracy. To ensure a comprehensive and robust evaluation of the sampling strategies, we select two widely used techniques:

\begin{itemize}
    \item \textbf{XGBoost~\cite{chen_XGBoostScalableTree_2016}:} XGBoost is a gradient-boosted decision tree model and highly efficient on structured, tabular data. It is a dominant method in many data science competitions and serves as a strong, non-deep-learning baseline.

    \item \textbf{DeepPerf~\cite{haDeepPerfPerformancePrediction2019}:} As a deep learning approach, DeepPerf is a specialized method that utilizes a deep feedforward neural network (FNN). 
    % A key feature of its design is the integration of sparsity regularization techniques, such as L1 regularization, which helps in identifying the most influential configuration options.
\end{itemize}

By employing both XGBoost and DeepPerf, we can assess whether the effectiveness of a given sampling strategy is universal or if it is dependent on the type of learning algorithm used. This model selection provides a more thorough validation of our sampling framework's capabilities.

\subsection{Evaluation Metrics}

\st{To ensure a fair comparison, we follow a standardized evaluation protocol. For a given sampling budget $k$, each sampling method selects $k$ configurations from the dataset. A performance model (e.g., XGBoost or DeepPerf) is then trained on the measured performance data of these $k$ samples. Subsequently, this trained model is used to predict the performance of all remaining, unseen configurations in the dataset. To ensure statistical significance, we repeat this entire process 5 times with different random seeds and report the average results.

We use the \textbf{Root Mean Square Error (RMSE)} to evaluate the quality of the generated performance models. RMSE measures the accuracy of the predictions by quantifying the magnitude of the error between the model's predicted values and the actual measured values. A lower RMSE indicates a more accurate model and, by extension, a more effective sampling strategy.}

\rv{Our framework, \tool{}, acts as a \textbf{configuration sampler}. Its goal is to select an informative subset of configurations for training a performance model. Therefore, to evaluate the effectiveness of our sampler, we follow standard practice and use a \textbf{downstream metric}: the prediction accuracy of a performance model trained on the samples selected by \tool{}.

Our evaluation protocol is as follows: for a given sampling budget $k$, each sampling method selects $k$ configurations. A performance model (e.g., XGBoost or DeepPerf) is then trained on the measured performance data of only these $k$ samples. Subsequently, this trained model is used to predict the performance of all remaining, unseen configurations in the dataset. To ensure statistical significance, we repeat this entire process 10 times with different random seeds and report the average results.

We use the \textbf{Root Mean Square Error (RMSE)} to evaluate the quality of these downstream performance models. RMSE measures the magnitude of the error between the model's predicted values and the actual measured values. A lower RMSE on the downstream model indicates that the selected $k$ samples were more informative and representative of the performance landscape. By extension, this signifies that the sampling strategy itself was more effective.}

\subsection{Studied LLMs and Implementation Details}
Our experiments include a diverse set of seven LLMs, encompassing both a closed-source model and various open-source alternatives. For the closed-source model, we use {\texttt{GPT-4o}}~\cite{opeai_gpt4}. The open-source models selected for our study are: the {\texttt{Qwen2.5}} series (\texttt{72B}, \texttt{32B}, and \texttt{7B})~\cite{qwen_qwen25_2025}, the {\texttt{DeepSeek}} series (\texttt{V3} and \texttt{R1})~\cite{deepseek-ai_deepseek-r1_2025}, \rv{the {\texttt{Mixtral}} series(\texttt{8\textbf{\(\times\)}7B} and \texttt{8\textbf{\(\times\)}22B})} and {\texttt{Llama3.3}}~\cite{grattafiori_llama_2024}.

The models are accessed via two different methods. The \texttt{Qwen2.5} series and \texttt{Llama3.3} are hosted and executed locally using the Ollama framework~\cite{_Ollama_}. For \texttt{GPT-4o}, \texttt{DeepSeek-V3}, and \texttt{DeepSeek-R1}, we utilized their official, publicly available APIs.

All experiments are conducted on a dedicated Linux server running Ubuntu 20.04.6 LTS. The server is equipped with an AMD 32-core processor, 1TB of RAM, and eight NVIDIA A6000 GPUs, ensuring a powerful and stable environment for our evaluations.

\section{Research Question}
\label{sec:research_question}

This section presents our empirical study investigating the effectiveness of using LLMs for software performance configuration sampling. Specifically, we employ our proposed framework, \textbf{\tool}, as an LLM-based sampler for the experiments. Our study aims to answer the following four research questions (RQs):

\begin{itemize}[leftmargin=*]
    \item \textbf{RQ1:} How does the effectiveness of \tool compare to baseline methods?
    
    \item \textbf{RQ2:} What are the key mechanisms through which an LLM contributes to \tool?

    % \item \textbf{RQ2:} How effective of \tool in pruning search space and feedback-based sampling?
    
    \item \textbf{RQ3:} How does LLMs selection impact \tool's effectiveness?
    
    \item \textbf{RQ4:} How do hyperparameters impact \tool?
\end{itemize}

The remainder of this section is organized around these research questions, with each subsection dedicated to one RQ.

\subsection*{RQ1: How does the effectiveness of \tool compare to baseline methods?}
\label{sec:rq1}
\begin{table*}[htbp]
\centering
\caption{Performance Comparison with Statistical Significance. Format for LLM4Perf: RMSE($\uparrow$Impr\%)*$^\wedge$L, where * indicates statistically significant(Wilcoxon p<0.05), $^\wedge$M/L indicate medium/large Cliff's delta effect size.}
\label{tab:full_performance_comparison}
\resizebox{\textwidth}{!}{%
\setlength{\tabcolsep}{6pt}
\renewcommand{\arraystretch}{1}
\begin{tabular}{l | c | c | l r r r r r r r r | l
r r r r r r r r }
\toprule
\multicolumn{3}{c|}{\textbf{Performance Modeling Method}} & \multicolumn{9}{c|}{\textbf{XGBoost}} & \multicolumn{9}{c}{\textbf{DeepPerf}} \\
\cmidrule(lr){1-21}
\makecell{\textbf{Project}\\\textbf{Name}} & \makecell{\textbf{Performance}\\\textbf{Metrics}} & \makecell{\textbf{Sample}\\\textbf{Size}} & \textbf{LLM4Perf} & \textbf{Random} & \textbf{Genetic} & \textbf{Flash} & \textbf{CoMSA} & \textbf{NSBS} & \textbf{EHVI} & \textbf{TSEMO} & \textbf{NSGA-III} & \textbf{LLM4Perf} & \textbf{Random} & \textbf{Genetic} & \textbf{Flash} & \textbf{CoMSA} & \textbf{NSBS} & \textbf{EHVI} & \textbf{TSEMO} & \textbf{NSGA-III} \\
\toprule
\multirow{14}{*}{LRZIP}
 & \multirow{7}{*}{\makecell{Compression\\Time}} & 10 & \cellcolor[HTML]{C6EFCE}{2.309($\uparrow$59.3\%)\textsuperscript{*L}} & 5.770 & 4.528 & 4.414 & 4.264 & 2.581 & 6.982 & 5.487 & 5.672 & \cellcolor[HTML]{C6EFCE}{3.283($\uparrow$49.8\%)\textsuperscript{*L}} & 6.388 & 6.323 & 7.171 & 7.238 & 8.686 & 7.654 & 7.393 & 6.537 \\
 &  & 20 & \cellcolor[HTML]{C6EFCE}{2.177($\uparrow$35.3\%)\textsuperscript{*L}} & 2.589 & 5.226 & 6.446 & 2.712 & 2.380 & 6.111 & 2.735 & 3.364 & \cellcolor[HTML]{C6EFCE}{2.235($\uparrow$55.7\%)\textsuperscript{*L}} & 5.188 & 6.633 & 6.052 & 6.303 & 3.444 & 6.588 & 6.698 & 5.045 \\
 &  & 30 & \cellcolor[HTML]{C6EFCE}{1.877($\uparrow$28.1\%)\textsuperscript{*L}} & 2.500 & 4.460 & 4.437 & 2.817 & 2.460 & 4.835 & 5.328 & 2.611 & \cellcolor[HTML]{C6EFCE}{2.182($\uparrow$66.4\%)\textsuperscript{*L}} & 3.693 & 19.101 & 8.227 & 5.766 & 5.170 & 6.280 & 4.411 & 6.489 \\
 &  & 40 & \cellcolor[HTML]{C6EFCE}{1.765($\uparrow$25.4\%)\textsuperscript{*L}} & 2.316 & 4.338 & 4.478 & 2.547 & 2.549 & 2.405 & 2.693 & 2.367 & \cellcolor[HTML]{C6EFCE}{1.882($\uparrow$37.0\%)\textsuperscript{*L}} & 3.364 & 6.447 & 6.017 & 4.265 & 3.124 & 4.405 & 3.873 & 2.988 \\
 &  & 50 & \cellcolor[HTML]{C6EFCE}{1.723($\uparrow$27.9\%)\textsuperscript{*L}} & 2.112 & 4.419 & 4.567 & 2.430 & 2.097 & 2.418 & 1.982 & 2.389 & \cellcolor[HTML]{C6EFCE}{1.919($\uparrow$47.7\%)\textsuperscript{*L}} & 2.685 & 7.511 & 5.480 & 3.931 & 2.785 & 2.755 & 3.530 & 3.668 \\
 &  & 60 & 1.784($\uparrow$34.2\%)\textsuperscript{*L} & 2.100 & 4.402 & 4.891 & 2.097 & 1.874 & 2.316 & \cellcolor[HTML]{C6EFCE}{1.776} & 2.710 & \cellcolor[HTML]{C6EFCE}{1.894($\uparrow$38.5\%)\textsuperscript{*L}} & 2.656 & 5.628 & 4.816 & 2.778 & 2.922 & 2.611 & 3.304 & 3.081 \\
 &  & 70 & 1.763($\uparrow$23.9\%)\textsuperscript{*L} & 1.821 & 4.412 & 4.956 & 1.931 & 1.765 & 2.192 & \cellcolor[HTML]{C6EFCE}{1.722} & 2.317 & \cellcolor[HTML]{C6EFCE}{1.628($\uparrow$46.8\%)\textsuperscript{*L}} & 2.768 & 4.996 & 5.519 & 3.073 & 2.065 & 2.550 & 2.544 & 3.062 \\
 \cmidrule(lr){2-21}
 & \multirow{7}{*}{\makecell{Max\\Memory}} & 10 & \cellcolor[HTML]{C6EFCE}{117.856($\uparrow$37.5\%)\textsuperscript{*L}} & 279.828 & 264.196 & 326.729 & 190.810 & 162.185 & 197.753 & 288.352 & 188.597 & \cellcolor[HTML]{C6EFCE}{160.422($\uparrow$64.9\%)\textsuperscript{*L}} & 299.967 & 285.665 & 315.273 & 248.333 & 399.796 & 368.336 & 327.333 & 456.856 \\
 &  & 20 & 105.116($\uparrow$23.4\%)\textsuperscript{*L} & 181.514 & 177.943 & 299.159 & 170.495 & \cellcolor[HTML]{C6EFCE}{102.590} & 195.747 & 150.312 & 137.246 & \cellcolor[HTML]{C6EFCE}{126.773($\uparrow$52.3\%)\textsuperscript{*L}} & 230.080 & 210.019 & 288.734 & 196.592 & 174.663 & 296.979 & 360.061 & 265.931 \\
 &  & 30 & \cellcolor[HTML]{C6EFCE}{92.652($\uparrow$32.2\%)\textsuperscript{*L}} & 120.716 & 199.463 & 284.815 & 114.817 & 100.141 & 112.555 & 166.306 & 136.620 & \cellcolor[HTML]{C6EFCE}{94.864($\uparrow$56.5\%)\textsuperscript{*L}} & 191.389 & 225.284 & 218.781 & 184.504 & 152.399 & 221.402 & 229.629 & 218.045 \\
 &  & 40 & \cellcolor[HTML]{C6EFCE}{97.595($\uparrow$26.9\%)\textsuperscript{M}} & 109.378 & 191.282 & 320.718 & 119.552 & 103.393 & 134.874 & 144.276 & 133.431 & \cellcolor[HTML]{C6EFCE}{106.165($\uparrow$36.7\%)\textsuperscript{*L}} & 169.603 & 221.970 & 265.254 & 175.293 & 132.043 & 208.348 & 149.503 & 167.741 \\
 &  & 50 & 99.461($\uparrow$4.8\%)\textsuperscript{M} & 108.745 & 177.727 & 270.428 & 116.405 & \cellcolor[HTML]{C6EFCE}{95.131} & 116.621 & 161.138 & 104.506 & \cellcolor[HTML]{C6EFCE}{88.361($\uparrow$64.1\%)\textsuperscript{*L}} & 163.084 & 219.045 & 256.488 & 158.584 & 156.608 & 174.234 & 151.047 & 245.933 \\
 &  & 60 & \cellcolor[HTML]{C6EFCE}{91.438($\uparrow$12.7\%)\textsuperscript{*L}} & 117.329 & 171.164 & 225.968 & 120.912 & 102.270 & 119.652 & 135.885 & 104.721 & \cellcolor[HTML]{C6EFCE}{88.714($\uparrow$72.0\%)\textsuperscript{*L}} & 159.078 & 188.610 & 291.621 & 156.404 & 176.371 & 190.114 & 152.315 & 316.903 \\
 &  & 70 & \cellcolor[HTML]{C6EFCE}{92.982($\uparrow$7.9\%)\textsuperscript{*L}} & 119.226 & 184.347 & 193.666 & 117.212 & 100.775 & 112.723 & 141.880 & 100.998 & \cellcolor[HTML]{C6EFCE}{91.421($\uparrow$56.9\%)\textsuperscript{*L}} & 135.551 & 194.090 & 200.673 & 173.460 & 204.199 & 175.899 & 141.932 & 212.316 \\
\midrule
\multirow{14}{*}{JavaGC}
 & \multirow{7}{*}{\makecell{GC\\Time}} & 10 & 0.366($\uparrow$18.2\%)\textsuperscript{M} & 0.364 & \cellcolor[HTML]{C6EFCE}{0.338} & 0.401 & 0.647 & 0.586 & 0.474 & 0.370 & 0.447 & \cellcolor[HTML]{C6EFCE}{0.211($\uparrow$52.1\%)\textsuperscript{*L}} & 0.272 & 0.212 & 0.251 & 0.423 & 0.964 & 0.237 & 0.233 & 0.441 \\
 &  & 20 & \cellcolor[HTML]{C6EFCE}{0.271($\uparrow$47.2\%)\textsuperscript{*L}} & 0.325 & 0.312 & 0.356 & 0.506 & 0.341 & 0.369 & 0.358 & 0.512 & \cellcolor[HTML]{C6EFCE}{0.200($\uparrow$34.7\%)\textsuperscript{*M}} & 0.226 & 0.267 & 0.350 & 0.373 & 0.282 & 0.212 & 0.207 & 0.306 \\
 &  & 30 & \cellcolor[HTML]{C6EFCE}{0.279($\uparrow$27.7\%)\textsuperscript{*L}} & 0.322 & 0.344 & 0.368 & 0.536 & 0.312 & 0.356 & 0.357 & 0.385 & 0.206($\uparrow$52.1\%)\textsuperscript{*L} & \cellcolor[HTML]{C6EFCE}{0.198} & 0.252 & 0.256 & 0.495 & 0.251 & 0.231 & 0.249 & 0.430 \\
 &  & 40 & 0.386($\downarrow$-9.9\%)\textsuperscript{} & \cellcolor[HTML]{C6EFCE}{0.319} & 0.325 & 0.330 & 0.472 & 0.321 & 0.369 & 0.337 & 0.351 & \cellcolor[HTML]{C6EFCE}{0.205($\uparrow$47.6\%)\textsuperscript{*M}} & 0.209 & 0.237 & 0.252 & 0.348 & 0.269 & 0.249 & 0.267 & 0.392 \\
 &  & 50 & 0.310($\downarrow$-5.9\%)\textsuperscript{} & \cellcolor[HTML]{C6EFCE}{0.293} & 0.295 & 0.379 & 0.576 & 0.328 & 0.387 & 0.329 & 0.293 & 0.230($\downarrow$7.4\%)\textsuperscript{} & 0.199 & \cellcolor[HTML]{C6EFCE}{0.196} & 0.385 & 0.468 & 0.226 & 0.247 & 0.211 & 0.215 \\
 &  & 60 & \cellcolor[HTML]{C6EFCE}{0.272($\uparrow$21.1\%)\textsuperscript{*L}} & 0.282 & 0.326 & 0.363 & 0.562 & 0.391 & 0.331 & 0.330 & 0.345 & 0.204($\uparrow$15.1\%)\textsuperscript{*L} & \cellcolor[HTML]{C6EFCE}{0.192} & 0.214 & 0.359 & 0.434 & 0.271 & 0.213 & 0.202 & 0.241 \\
 &  & 70 & \cellcolor[HTML]{C6EFCE}{0.293($\uparrow$2.9\%)} & 0.317 & 0.294 & 0.373 & 0.548 & 0.298 & 0.329 & 0.333 & 0.302 & 0.217($\uparrow$5.1\%)\textsuperscript{M} & \cellcolor[HTML]{C6EFCE}{0.194} & 0.194 & 0.314 & 0.355 & 0.218 & 0.230 & 0.233 & 0.228 \\
 \cmidrule(lr){2-21}
 & \multirow{7}{*}{\makecell{Average\\Pause\\Time}} & 10 & \cellcolor[HTML]{C6EFCE}{2.458($\uparrow$36.5\%)\textsuperscript{*L}} & 4.119 & 3.850 & 3.855 & 3.744 & 13.970 & 4.537 & 4.187 & 3.871 & \cellcolor[HTML]{C6EFCE}{2.058($\uparrow$36.3\%)\textsuperscript{*L}} & 3.340 & 3.578 & 3.057 & 3.744 & 16.611 & 5.322 & 3.419 & 3.229 \\
 &  & 20 & \cellcolor[HTML]{C6EFCE}{2.459($\uparrow$40.8\%)\textsuperscript{*L}} & 3.872 & 3.601 & 4.069 & 3.885 & 2.778 & 4.530 & 4.217 & 4.157 & \cellcolor[HTML]{C6EFCE}{2.131($\uparrow$42.7\%)\textsuperscript{*L}} & 2.827 & 4.076 & 5.161 & 3.766 & 2.689 & 3.200 & 3.467 & 3.719 \\
 &  & 30 & 2.744($\uparrow$30.8\%)\textsuperscript{*L} & 3.752 & 3.824 & 3.921 & 3.800 & \cellcolor[HTML]{C6EFCE}{2.613} & 4.014 & 3.860 & 3.963 & \cellcolor[HTML]{C6EFCE}{2.071($\uparrow$45.7\%)\textsuperscript{*M}} & 3.718 & 3.261 & 2.974 & 3.347 & 2.663 & 3.089 & 2.770 & 3.811 \\
 &  & 40 & \cellcolor[HTML]{C6EFCE}{2.525($\uparrow$37.7\%)\textsuperscript{*L}} & 3.395 & 3.599 & 3.830 & 4.015 & 3.052 & 3.946 & 3.825 & 4.050 & \cellcolor[HTML]{C6EFCE}{2.257($\uparrow$24.6\%)\textsuperscript{*L}} & 2.519 & 3.233 & 5.607 & 3.579 & 3.403 & 2.994 & 3.008 & 2.992 \\
 &  & 50 & 3.060($\uparrow$16.4\%)\textsuperscript{*L} & 3.294 & 3.782 & 3.712 & 3.908 & \cellcolor[HTML]{C6EFCE}{3.021} & 3.537 & 3.908 & 3.659 & \cellcolor[HTML]{C6EFCE}{1.925($\uparrow$37.7\%)\textsuperscript{M}} & 3.274 & 3.018 & 2.628 & 3.062 & 2.950 & 3.225 & 2.992 & 3.089 \\
 &  & 60 & \cellcolor[HTML]{C6EFCE}{2.665($\uparrow$37.9\%)\textsuperscript{*L}} & 3.401 & 3.802 & 3.950 & 3.908 & 2.786 & 3.988 & 3.713 & 4.291 & \cellcolor[HTML]{C6EFCE}{1.891($\uparrow$30.8\%)\textsuperscript{M}} & 2.458 & 2.517 & 2.638 & 3.939 & 2.512 & 3.021 & 2.290 & 2.732 \\
 &  & 70 & 2.712($\uparrow$28.2\%)\textsuperscript{*L} & 2.816 & 3.036 & 3.950 & 3.808 & \cellcolor[HTML]{C6EFCE}{2.618} & 3.743 & 3.722 & 3.778 & \cellcolor[HTML]{C6EFCE}{2.088($\uparrow$31.1\%)\textsuperscript{*L}} & 3.088 & 2.629 & 3.049 & 2.971 & 2.374 & 2.857 & 2.714 & 3.031 \\
\midrule
\multirow{14}{*}{SQLite}
 & \multirow{7}{*}{\makecell{Response\\Time}} & 10 & \cellcolor[HTML]{C6EFCE}{0.050($\uparrow$68.6\%)\textsuperscript{M}} & 0.515 & 0.147 & 0.169 & 0.178 & 0.523 & 0.148 & 0.057 & 0.160 & 0.106($\downarrow$1.7\%)\textsuperscript{} & 0.286 & 0.523 & 0.316 & 0.126 & 0.288 & 0.222 & 0.314 & \cellcolor[HTML]{C6EFCE}{0.105} \\
 &  & 20 & 0.049($\uparrow$68.1\%)\textsuperscript{M} & 0.057 & 0.146 & 0.153 & 0.049 & 0.056 & 0.059 & \cellcolor[HTML]{C6EFCE}{0.048} & 0.153 & \cellcolor[HTML]{C6EFCE}{0.058($\uparrow$48.0\%)\textsuperscript{M}} & 0.082 & 0.231 & 0.127 & 0.202 & 0.064 & 0.075 & 0.064 & 0.112 \\
 &  & 30 & \cellcolor[HTML]{C6EFCE}{0.040($\uparrow$74.3\%)\textsuperscript{*L}} & 0.045 & 0.057 & 0.140 & 0.055 & 0.052 & 0.056 & 0.047 & 0.154 & \cellcolor[HTML]{C6EFCE}{0.031($\uparrow$65.6\%)\textsuperscript{*L}} & 0.055 & 0.178 & 0.140 & 0.046 & 0.041 & 0.114 & 0.053 & 0.091 \\
 &  & 40 & 0.045($\uparrow$16.1\%)\textsuperscript{*M} & 0.049 & 0.160 & 0.151 & \cellcolor[HTML]{C6EFCE}{0.044} & 0.047 & 0.056 & 0.048 & 0.054 & 0.070($\downarrow$98.0\%)\textsuperscript{} & 0.042 & 0.132 & 0.135 & 0.055 & \cellcolor[HTML]{C6EFCE}{0.034} & 0.065 & 0.042 & 0.035 \\
 &  & 50 & 0.045($\uparrow$18.8\%)\textsuperscript{*M} & \cellcolor[HTML]{C6EFCE}{0.043} & 0.160 & 0.155 & 0.046 & 0.044 & 0.052 & 0.046 & 0.056 & \cellcolor[HTML]{C6EFCE}{0.034($\uparrow$54.6\%)\textsuperscript{*L}} & 0.042 & 0.140 & 0.146 & 0.113 & 0.034 & 0.113 & 0.037 & 0.075 \\
 &  & 60 & \cellcolor[HTML]{C6EFCE}{0.041($\uparrow$21.6\%)\textsuperscript{*L}} & 0.048 & 0.180 & 0.143 & 0.048 & 0.044 & 0.054 & 0.046 & 0.052 & 0.035($\uparrow$18.4\%)\textsuperscript{*M} & 0.037 & 0.131 & 0.165 & 0.115 & 0.034 & 0.055 & \cellcolor[HTML]{C6EFCE}{0.032} & 0.043 \\
 &  & 70 & \cellcolor[HTML]{C6EFCE}{0.042($\uparrow$19.8\%)\textsuperscript{*L}} & 0.043 & 0.152 & 0.153 & 0.044 & 0.042 & 0.049 & 0.045 & 0.052 & \cellcolor[HTML]{C6EFCE}{0.031($\uparrow$43.0\%)\textsuperscript{*L}} & 0.040 & 0.097 & 0.152 & 0.038 & 0.032 & 0.118 & 0.044 & 0.055 \\
 \cmidrule(lr){2-21}
 & \multirow{7}{*}{\makecell{Max\\Memory}} & 10 & \cellcolor[HTML]{C6EFCE}{1.919($\uparrow$21.7\%)\textsuperscript{*L}} & 2.816 & 2.563 & 2.967 & 2.564 & 2.446 & 2.701 & 2.628 & 2.451 & \cellcolor[HTML]{C6EFCE}{58.712($\uparrow$29.6\%)\textsuperscript{M}} & 101.152 & 79.825 & 74.043 & 88.957 & 106.549 & 60.054 & 85.189 & 83.448 \\
 &  & 20 & \cellcolor[HTML]{C6EFCE}{1.918($\uparrow$28.5\%)\textsuperscript{*L}} & 2.465 & 2.690 & 2.930 & 2.403 & 2.563 & 2.698 & 2.729 & 2.682 & \cellcolor[HTML]{C6EFCE}{2.875($\uparrow$95.5\%)\textsuperscript{*M}} & 42.499 & 52.900 & 83.112 & 64.812 & 46.484 & 76.365 & 44.621 & 64.353 \\
 &  & 30 & \cellcolor[HTML]{C6EFCE}{1.918($\uparrow$32.3\%)\textsuperscript{*L}} & 2.495 & 2.451 & 3.843 & 2.352 & 2.411 & 2.525 & 2.576 & 2.833 & 27.425($\uparrow$65.7\%)\textsuperscript{*L} & 35.441 & 76.566 & 55.435 & \cellcolor[HTML]{C6EFCE}{20.926} & 24.680 & 23.488 & 29.221 & 79.941 \\
 &  & 40 & \cellcolor[HTML]{C6EFCE}{1.917($\uparrow$22.4\%)\textsuperscript{*L}} & 2.558 & 2.323 & 4.429 & 2.671 & 2.512 & 2.439 & 2.493 & 2.469 & 2.360($\uparrow$88.4\%)\textsuperscript{M} & 3.064 & 2.145 & 48.467 & 44.780 & 23.539 & 100.997 & \cellcolor[HTML]{C6EFCE}{2.076} & 20.361 \\
 &  & 50 & \cellcolor[HTML]{C6EFCE}{1.918($\uparrow$27.2\%)\textsuperscript{*L}} & 2.399 & 2.569 & 4.676 & 2.586 & 2.323 & 2.489 & 2.293 & 2.637 & 8.280($\uparrow$56.5\%)\textsuperscript{M} & \cellcolor[HTML]{C6EFCE}{2.145} & 38.220 & 39.355 & 6.856 & 2.212 & 2.411 & 11.505 & 19.036 \\
 &  & 60 & \cellcolor[HTML]{C6EFCE}{1.918($\uparrow$29.6\%)\textsuperscript{*L}} & 2.610 & 2.362 & 3.921 & 2.657 & 2.376 & 2.426 & 2.436 & 2.724 & \cellcolor[HTML]{C6EFCE}{1.892($\uparrow$96.1\%)\textsuperscript{*L}} & 6.137 & 30.455 & 45.462 & 2.467 & 2.732 & 11.618 & 2.072 & 48.029 \\
 &  & 70 & \cellcolor[HTML]{C6EFCE}{1.917($\uparrow$24.2\%)\textsuperscript{*L}} & 2.375 & 2.552 & 4.355 & 2.619 & 2.394 & 2.474 & 2.342 & 2.529 & \cellcolor[HTML]{C6EFCE}{2.134($\uparrow$76.2\%)\textsuperscript{M}} & 2.840 & 29.304 & 33.118 & 5.708 & 4.063 & 6.324 & 3.896 & 8.962 \\
\midrule
\multirow{14}{*}{X264}
 & \multirow{7}{*}{\makecell{Encoding\\Time}} & 10 & \cellcolor[HTML]{C6EFCE}{6.214($\uparrow$85.1\%)\textsuperscript{*L}} & 14.304 & 9.655 & 39.572 & 17.284 & 9.106 & 45.765 & 22.678 & 41.769 & \cellcolor[HTML]{C6EFCE}{4.363($\uparrow$87.0\%)\textsuperscript{*M}} & 31.321 & 26.292 & 33.141 & 31.829 & 28.399 & 38.815 & 33.315 & 33.538 \\
 &  & 20 & \cellcolor[HTML]{C6EFCE}{5.849($\uparrow$85.4\%)\textsuperscript{*L}} & 7.556 & 7.022 & 39.002 & 12.847 & 7.384 & 39.563 & 22.684 & 39.967 & \cellcolor[HTML]{C6EFCE}{4.055($\uparrow$83.6\%)\textsuperscript{*L}} & 11.211 & 16.415 & 36.251 & 11.864 & 5.101 & 20.084 & 23.926 & 24.721 \\
 &  & 30 & \cellcolor[HTML]{C6EFCE}{4.160($\uparrow$84.6\%)\textsuperscript{*L}} & 5.951 & 6.685 & 39.275 & 6.932 & 6.119 & 26.035 & 23.268 & 27.023 & \cellcolor[HTML]{C6EFCE}{3.281($\uparrow$35.6\%)\textsuperscript{*M}} & 10.684 & 6.400 & 22.261 & 3.736 & 4.278 & 12.896 & 20.878 & 5.093 \\
 &  & 40 & 4.639($\uparrow$88.1\%)\textsuperscript{*L} & 4.950 & 5.839 & 31.472 & 5.752 & \cellcolor[HTML]{C6EFCE}{4.393} & 23.990 & 22.469 & 39.058 & \cellcolor[HTML]{C6EFCE}{3.662($\uparrow$80.1\%)\textsuperscript{*L}} & 7.111 & 5.266 & 15.941 & 5.414 & 4.072 & 11.428 & 8.006 & 18.390 \\
 &  & 50 & 4.354($\uparrow$1.9\%) & 4.228 & 4.747 & 31.053 & 5.922 & \cellcolor[HTML]{C6EFCE}{4.084} & 23.046 & 22.485 & 4.440 & 3.100($\uparrow$48.6\%)\textsuperscript{*L} & \cellcolor[HTML]{C6EFCE}{3.083} & 4.592 & 34.279 & 4.613 & 4.574 & 7.166 & 16.673 & 6.035 \\
 &  & 60 & 4.558($\uparrow$42.7\%)\textsuperscript{*L} & 4.806 & 4.889 & 31.663 & 4.934 & \cellcolor[HTML]{C6EFCE}{4.183} & 23.093 & 23.295 & 7.959 & 2.932($\uparrow$42.5\%)\textsuperscript{*L} & 2.695 & 3.449 & 16.717 & 3.863 & \cellcolor[HTML]{C6EFCE}{2.679} & 18.599 & 6.237 & 5.097 \\
 &  & 70 & \cellcolor[HTML]{C6EFCE}{3.635($\uparrow$51.1\%)\textsuperscript{*L}} & 3.969 & 4.575 & 18.694 & 4.934 & 4.048 & 23.363 & 23.300 & 7.429 & 3.057($\uparrow$34.5\%)\textsuperscript{*L} & \cellcolor[HTML]{C6EFCE}{2.935} & 10.267 & 18.398 & 3.631 & 2.936 & 9.497 & 12.383 & 4.663 \\
 \cmidrule(lr){2-21}
 & \multirow{7}{*}{\makecell{PSNR}} & 10 & \cellcolor[HTML]{C6EFCE}{0.551($\uparrow$34.3\%)\textsuperscript{*L}} & 0.697 & 0.868 & 1.468 & 1.065 & 0.827 & 0.909 & 0.749 & 0.839 & 0.478($\uparrow$11.5\%)\textsuperscript{M} & 0.490 & 5.727 & 0.871 & 0.591 & \cellcolor[HTML]{C6EFCE}{0.475} & 10.298 & 11.824 & 0.540 \\
 &  & 20 & \cellcolor[HTML]{C6EFCE}{0.493($\uparrow$39.1\%)\textsuperscript{*L}} & 0.770 & 0.878 & 1.444 & 0.961 & 0.661 & 0.714 & 0.776 & 0.808 & \cellcolor[HTML]{C6EFCE}{0.458($\uparrow$42.4\%)\textsuperscript{*L}} & 0.638 & 1.312 & 3.286 & 0.507 & 0.466 & 0.476 & 0.487 & 0.794 \\
 &  & 30 & \cellcolor[HTML]{C6EFCE}{0.490($\uparrow$39.7\%)\textsuperscript{*L}} & 0.663 & 0.709 & 1.292 & 0.903 & 0.676 & 0.706 & 0.771 & 0.814 & 0.458($\uparrow$85.1\%)\textsuperscript{*L} & \cellcolor[HTML]{C6EFCE}{0.443} & 0.496 & 0.607 & 1.716 & 0.473 & 0.454 & 0.450 & 3.069 \\
 &  & 40 & \cellcolor[HTML]{C6EFCE}{0.490($\uparrow$36.0\%)\textsuperscript{*L}} & 0.679 & 0.776 & 1.089 & 0.852 & 0.628 & 0.714 & 0.665 & 0.765 & 0.471($\downarrow$1.4\%)\textsuperscript{} & \cellcolor[HTML]{C6EFCE}{0.432} & 1.336 & 0.816 & 0.606 & 0.570 & 0.435 & 0.449 & 0.464 \\
 &  & 50 & \cellcolor[HTML]{C6EFCE}{0.491($\uparrow$40.3\%)\textsuperscript{*L}} & 0.687 & 0.724 & 1.172 & 0.887 & 0.662 & 0.721 & 0.778 & 0.823 & 0.448($\uparrow$39.7\%)\textsuperscript{*L} & 0.427 & 1.380 & 0.771 & 0.860 & \cellcolor[HTML]{C6EFCE}{0.424} & 0.434 & 1.292 & 0.744 \\
 &  & 60 & \cellcolor[HTML]{C6EFCE}{0.489($\uparrow$35.9\%)\textsuperscript{*L}} & 0.653 & 0.789 & 1.079 & 0.819 & 0.623 & 0.717 & 0.659 & 0.764 & 0.452($\downarrow$1.6\%)\textsuperscript{} & \cellcolor[HTML]{C6EFCE}{0.427} & 0.430 & 0.472 & 0.729 & 0.455 & 0.621 & 0.611 & 0.445 \\
 &  & 70 & \cellcolor[HTML]{C6EFCE}{0.490($\uparrow$33.6\%)\textsuperscript{*L}} & 0.620 & 0.690 & 1.228 & 0.794 & 0.600 & 0.699 & 0.660 & 0.737 & 0.448($\uparrow$59.3\%)\textsuperscript{*L} & 0.426 & 0.471 & 0.738 & 0.426 & \cellcolor[HTML]{C6EFCE}{0.425} & 1.431 & 0.716 & 1.101 \\
 \bottomrule
\end{tabular}
}
\end{table*}

\begin{figure*}[tbp]
    \centering
    \includegraphics[width=\textwidth]{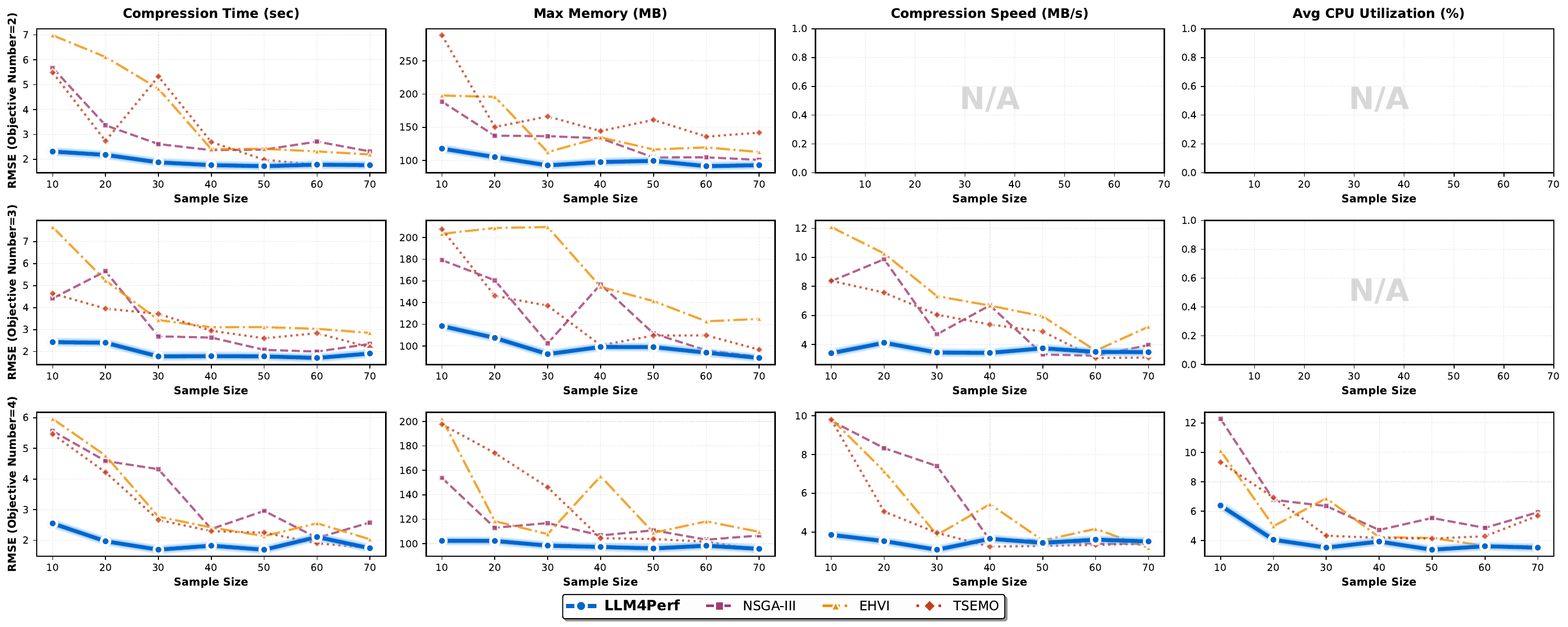}
    \caption{Scalability comparison (RMSE) of \tool (blue line) against multi-objective baselines (NSGA-III, EHVI, TSEMO) on LRZIP, varying the number of target objectives (2, 3, and 4). Lower RMSE is better. ``N/A'' marks metrics not included.}
    \label{fig:different_metrics}
\end{figure*}

\noindent \textbf{\textit{Motivation.}} \rv{Having introduced the framework, this section evaluates the effectiveness of \textbf{\tool}. We design our evaluation to assess three key perspectives:} \begin{enumerate*}[label=(\roman*)] \item its overall prediction accuracy compared to baseline methods, \item its generalizability across different performance modeling techniques, and \item its capability in handling multi-objective performance scenarios. \end{enumerate*}

\noindent  \textbf{\textit{Approach.}}

To evaluate our framework, \textbf{\tool}, we compare it against eight representative baseline methods on four software projects.

\rv{Our evaluation is designed to assess two key aspects: \begin{itemize} \item \rv{\textbf{Project Generality:} We first compare \textbf{\tool} against all baselines on the four projects, each involving two distinct performance metrics (results in Table~\ref{tab:full_performance_comparison}).} \item \rv{\textbf{Scalability with Objectives:} We further extend the evaluation to four performance metrics, as shown in Figure~\ref{fig:different_metrics}.} \end{itemize}}

For all experiments, we use the configurations selected by each sampler to train two distinct performance models, XGBoost and DeepPerf. The effectiveness is measured by the downstream Root Mean Square Error (RMSE) of these models. To ensure statistical robustness, we test across seven sampling budgets (from 10 to 70) and repeat each experiment ten times. \rv{To validate our findings, we conduct a non-parametric statistical analysis using the Wilcoxon signed-rank test (p < 0.05) and calculate the Cliff's delta effect size ($\delta \ge 0.33$) to measure practical significance. Specifically, this statistical analysis is performed for the comparison between \tool and NSGA-III, which serves as a widely adopted multi-objective optimization baseline~\cite{yuan_ImprovedNSGAIIIProcedure_2014, mkaouer_ManyObjectiveSoftwareRemodularization_2015, ishibuchi_PerformanceComparisonNSGAII_2016,cui_ImprovedNSGAIIISelectionandelimination_2019}.} It is important to note that the single-objective baselines (e.g., Random, Genetic) must be run independently for each performance metric, whereas \textbf{\tool} and the multi-objective baselines optimize all metrics in a single run.

\noindent \textit{\textbf{Results.}} \textbf{\tool generally achieves higher prediction accuracy compared to baseline methods.} Across the 112 evaluation scenarios (4 systems $\times$ 2 metrics $\times$ 7 sample sizes $\times$ 2 models), Table~\ref{tab:full_performance_comparison} shows that \tool achieves the lowest RMSE in \textbf{\rv{77} cases}, representing \textbf{\rv{68.8\%}} of all scenarios. This demonstrates the clear superiority of \tool. While a strong baseline like \texttt{NSBS} may occasionally perform slightly better (e.g., X264 Encoding Time, XGBoost, size 40), the difference is often marginal. In contrast, when \tool outperforms other methods, its advantage can be substantial. For instance, on the \texttt{SQLite} \textit{Max Memory} metric (DeepPerf, size 30), \tool's RMSE is 27.425, whereas \sout{\texttt{NSGA-II} yields an error of 76.566} \rv{\texttt{NSGA-III} yields an error of 79.941}, highlighting \tool's ability to avoid poor sampling choices.

\textbf{\tool demonstrates superior multi-objective handling over classic \rv{and modern} algorithms.}The framework's strength in balancing conflicting metrics is illustrated by its comparison with \st{\texttt{NSGA-II}} \rv{algorithms designed specifically for multi-objective optimization, such as \texttt{NSGA-III} and \texttt{EHVI}}. \st{On every multi-metric benchmark, LLM4Perf consistently provides samples that lead to more accurate models.} On the \texttt{X264} benchmark, which has a strong trade-off between \textit{Encoding Time} and \textit{Video Quality (PSNR)}, \tool's superiority is clear. For example, with 70 samples and a DeepPerf model, \tool achieves an RMSE of \st{2.470} \rv{3.057} for \textit{Encoding Time} and \st{0.443} \rv{0.448} for \textit{PSNR}, while \st{NSGA-II's corresponding errors are 5.446 and 0.461} \rv{\texttt{NSGA-III}'s corresponding errors are 4.663 and 1.101, and \texttt{EHVI}'s are 9.497 and 1.431}. This \rv{shows \tool is more accurate on both conflicting metrics}, suggesting its reasoning about semantic trade-offs is more effective than traditional evolutionary search.

\textbf{\tool's sampling strategy is effective across different performance modeling techniques.}The effectiveness of \tool's sampling is not dependent on the final performance model used. This is clearly demonstrated across all systems. For instance, on the \texttt{LRZIP} benchmark for \textit{Compression Time} (sample size 40), samples selected by \tool yield the best-performing model for both the tree-based \texttt{XGBoost} (RMSE of \sout{1.684} \rv{1.765}) and the neural network-based \texttt{DeepPerf} (RMSE of \sout{1.899} \rv{1.882}). \rv{In both cases, these results are superior to all eight other baselines.} This finding confirms that \tool identifies highly representative configurations that generalize well across different performance models.

\rv{\textbf{The observed improvements are statistically and practically significant.} As detailed in Table~\ref{tab:full_performance_comparison}, the analysis confirms that \tool's superiority is not due to chance. The performance improvement is statistically significant (marked with *) in \textbf{89 of the 112} scenarios (79.5\%). Moreover, the magnitude of this improvement is practically meaningful: a medium (M) or large (L) effect size was observed in \textbf{103 of the 112} scenarios (91.96\%). Notably, \textbf{80} of these cases demonstrated a large effect size (L), highlighting that \tool provides a substantial and consistent advantage over the baseline methods.}

\rv{
\noindent \textbf{\tool demonstrates superior scalability as the number of objectives increases.} Figure~\ref{fig:different_metrics} clearly illustrates how \tool's performance scales as the number of target objectives increases, a scenario where multi-objective baselines often falter. In the \textit{objective Number=2} scenario, \tool (blue line) already establishes a clear lead, consistently achieving the lowest RMSE for both \textit{Compression Time} and \textit{Max Memory}. As the complexity scales to \textit{objective Number=3} (adding \textit{Compression Speed}) and \textit{objective Number=4} (adding \textit{Avg CPU Utilization}), the performance gap between \tool and the baselines widens. While \tool's RMSE remains low and stable, the error rates for the baselines degrade, becoming volatile and worse.

\noindent \textbf{The baselines' effectiveness diminishes in high-dimensional objective spaces.} This degradation is most pronounced in the \textit{Objective Number=4} (XGBoost) scenarios. For instance, when modeling \textit{Avg CPU Utilization} with XGBoost, \tool maintains a low and stable RMSE (around 4-6). In contrast, \texttt{TSEMO}'s error starts high (around 9) and remains higher, while \texttt{EHVI}'s error also fluctuates consistently above \tool's. A similar trend is visible for \textit{Max Memory} (\textit{Objective Number=4}, XGBoost), where \tool's RMSE hovers around 100, while \texttt{EHVI}'s error starts above 200, and \texttt{NSGA-III}'s error also remains high (above 140). This strongly suggests that as the complexity of performance trade-offs increases, the ability of traditional algorithms to find representative samples diminishes. Conversely, \tool's LLM-based reasoning effectively navigates these complex, high-dimensional objective spaces, consistently identifying superior configurations even with small sample budgets.
}

\greybox{\textbf{RQ1 Summary:} \tool demonstrates superior performance across both \textbf{project generality} and \textbf{objective scalability}. First, when compared against eight baselines on four distinct projects (Table~\ref{tab:full_performance_comparison}), \tool achieves the lowest RMSE in \textbf{68.8\% (77/112)} of all scenarios, and this effectiveness is consistent across both XGBoost and DeepPerf models. Second, as the number of objectives increases from two to four (Figure~\ref{fig:different_metrics}), \tool's performance remains high and stable. In contrast, the effectiveness of specialized multi-objective baselines degrades in high-dimensional objective spaces.}

\subsection*{RQ2: What are the key mechanisms through which an LLM contributes to \tool?}
\label{sec:rq2}

\noindent \textbf{\textit{Motivation.}} The results from RQ1 establish the effectiveness of \textbf{\tool}. This section delves deeper to understand the underlying reasons for its effectiveness. Our analysis therefore investigates \textbf{the two core mechanisms} behind \tool's effectiveness: its ability to perform \begin{enumerate*}[label=(\roman*)] \item \textit{space pruning} by filtering insensitive configurations, and \item \textit{sampling strategy refinement based on iterative feedback.} \end{enumerate*} By investigating these two mechanisms, we aim to provide insights into how LLMs can be leveraged for effective configuration sampling.

\noindent \textbf{\textit{Approach.}} To analyze these two mechanisms, we design two distinct experiments. First, to assess the impact of \textbf{space pruning}, we conduct an experiment to determine if baseline methods also benefit from a pruned configuration space. For each system, we use \tool to generate a pruned configuration space by removing configurations it deems insensitive. We then apply all baseline samplers to both the full space and this new pruned space, comparing the resulting RMSE of the trained performance models. To ensure a fair comparison, all models are evaluated on the full space configuration space. Second, to illustrate the effectiveness of \textbf{adaptive strategy refinement}, we conduct a qualitative analysis of the \tool's iterative feedback loop. We refer back to the \texttt{LRZIP} \textbf{Running Example} (presented in Section 3) to analyze how the \textbf{Performance Trend Analyzer} and \textbf{Sampling Strategy Designer} use observed data to progressively refine the sampling strategy and focus on performance-critical regions.

\begin{table*}[htbp]
\centering
\caption{Impact of Filtering Insensitive Options on Model Performance (RMSE). Lower is better.}
\label{tab:filtering_effectiveness}

% 使用 resizebox 將極寬的表格縮放到文字寬度
\resizebox{\textwidth}{!}{%
% 調整行高讓內容更清晰
\renewcommand{\arraystretch}{0.5}
% 調整列間距
\setlength{\tabcolsep}{12pt}
\begin{tabular}{ll | rr | rr | rr | rr | rr | rr | rr | rr}
\toprule
\multirow{3}{*}{\textbf{Method}} & \multirow{3}{*}{\makecell[c]{\textbf{Sample}\\\textbf{Size}}} & \multicolumn{4}{c}{\textbf{LRZIP} \textit{(Compression Time)}} & \multicolumn{4}{c}{\textbf{JavaGC} \textit{(GC Time)}} & \multicolumn{4}{c}{\textbf{SQLite} \textit{(Response Time)}} & \multicolumn{4}{c}{\textbf{X264} \textit{(Encoding Time)}} \\

\cmidrule(lr){3-6} \cmidrule(lr){7-10} \cmidrule(lr){11-14} \cmidrule(lr){15-18}
& & \multicolumn{2}{c}{\textbf{XGBoost}} & \multicolumn{2}{c}{\textbf{DeepPerf}} & \multicolumn{2}{c}{\textbf{XGBoost}} & \multicolumn{2}{c}{\textbf{DeepPerf}} & \multicolumn{2}{c}{\textbf{XGBoost}} & \multicolumn{2}{c}{\textbf{DeepPerf}} & \multicolumn{2}{c}{\textbf{XGBoost}} & \multicolumn{2}{c}{\textbf{DeepPerf}} \\
\cmidrule(lr){3-4} \cmidrule(lr){5-6} \cmidrule(lr){7-8} \cmidrule(lr){9-10} \cmidrule(lr){11-12} \cmidrule(lr){13-14} \cmidrule(lr){15-16} \cmidrule(lr){17-18}
& & \textbf{Full} & \textbf{Pruned} & \textbf{Full} & \textbf{Pruned} & \textbf{Full} & \textbf{Pruned} & \textbf{Full} & \textbf{Pruned} & \textbf{Full} & \textbf{Pruned} & \textbf{Full} & \textbf{Pruned} & \textbf{Full} & \textbf{Pruned} & \textbf{Full} & \textbf{Pruned} \\
\midrule % 正式的数据行从此开始
\multirow{7}{*}{Random}
& 10 & 5.770 & \cellcolor[HTML]{C6EFCE}{4.174} & 6.388 & \cellcolor[HTML]{C6EFCE}{4.119} & 0.364 & \cellcolor[HTML]{C6EFCE}{0.302} & 0.272 & \cellcolor[HTML]{C6EFCE}{0.236} & 0.515 & \cellcolor[HTML]{C6EFCE}{0.067} & 0.286 & \cellcolor[HTML]{C6EFCE}{0.139} & 14.304 & 19.843 & 31.321 & \cellcolor[HTML]{C6EFCE}{14.144}\\
& 20 & 2.589 & \cellcolor[HTML]{C6EFCE}{2.383} & 5.188 & \cellcolor[HTML]{C6EFCE}{2.599} & 0.325 & \cellcolor[HTML]{C6EFCE}{0.323} & 0.226 & \cellcolor[HTML]{C6EFCE}{0.207} & 0.057 & \cellcolor[HTML]{C6EFCE}{0.049} & 0.082 & \cellcolor[HTML]{C6EFCE}{0.055} & 7.556 & \cellcolor[HTML]{C6EFCE}{5.778} & 11.211 & \cellcolor[HTML]{C6EFCE}{4.925}\\
& 30 & 2.500 & \cellcolor[HTML]{C6EFCE}{2.150} & 3.693 & \cellcolor[HTML]{C6EFCE}{2.618} & 0.322 & 0.360 & 0.198 & 0.202 & 0.045 & \cellcolor[HTML]{C6EFCE}{0.042} & 0.055 & \cellcolor[HTML]{C6EFCE}{0.039} & 5.951 & \cellcolor[HTML]{C6EFCE}{5.227} & 10.684 & \cellcolor[HTML]{C6EFCE}{3.711}\\
& 40 & 2.316 & \cellcolor[HTML]{C6EFCE}{1.837} & 3.364 & \cellcolor[HTML]{C6EFCE}{2.118} & 0.319 & \cellcolor[HTML]{C6EFCE}{0.281} & 0.209 & 0.235 & 0.049 & \cellcolor[HTML]{C6EFCE}{0.043} & 0.042 & \cellcolor[HTML]{C6EFCE}{0.032} & 4.950 & \cellcolor[HTML]{C6EFCE}{4.518} & 7.111 & \cellcolor[HTML]{C6EFCE}{3.382}\\
& 50 & 2.112 & \cellcolor[HTML]{C6EFCE}{1.770} & 2.685 & \cellcolor[HTML]{C6EFCE}{2.203} & 0.293 & \cellcolor[HTML]{C6EFCE}{0.271} & 0.199 & 0.199 & 0.043 & 0.043 & 0.042 & \cellcolor[HTML]{C6EFCE}{0.032} & 4.228 & \cellcolor[HTML]{C6EFCE}{3.963} & 3.083 & \cellcolor[HTML]{C6EFCE}{2.420}\\
& 60 & 2.100 & \cellcolor[HTML]{C6EFCE}{1.839} & 2.656 & \cellcolor[HTML]{C6EFCE}{2.212} & 0.282 & 0.289 & 0.192 & \cellcolor[HTML]{C6EFCE}{0.189} & 0.048 & \cellcolor[HTML]{C6EFCE}{0.041} & 0.037 & \cellcolor[HTML]{C6EFCE}{0.033} & 4.806 & \cellcolor[HTML]{C6EFCE}{3.505} & 2.695 & \cellcolor[HTML]{C6EFCE}{2.681}\\
& 70 & 1.821 & 1.866 & 2.768 & \cellcolor[HTML]{C6EFCE}{2.021} & 0.317 & \cellcolor[HTML]{C6EFCE}{0.295} & 0.194 & \cellcolor[HTML]{C6EFCE}{0.192} & 0.043 & 0.044 & 0.040 & \cellcolor[HTML]{C6EFCE}{0.032} & 3.969 & \cellcolor[HTML]{C6EFCE}{3.231} & 2.935 & \cellcolor[HTML]{C6EFCE}{2.327}\\
\midrule
\multirow{7}{*}{Genetic}
& 10 & 4.528 & \cellcolor[HTML]{C6EFCE}{3.477} & 6.323 & \cellcolor[HTML]{C6EFCE}{3.158} & 0.325 & \cellcolor[HTML]{C6EFCE}{0.324} & 0.220 & \cellcolor[HTML]{C6EFCE}{0.206} & 0.147 & \cellcolor[HTML]{C6EFCE}{0.056} & 0.523 & \cellcolor[HTML]{C6EFCE}{0.078} & 16.882 & \cellcolor[HTML]{C6EFCE}{10.387} & 27.133 & \cellcolor[HTML]{C6EFCE}{25.543}\\
& 20 & 5.226 & \cellcolor[HTML]{C6EFCE}{3.525} & 6.633 & \cellcolor[HTML]{C6EFCE}{3.854} & 0.325 & \cellcolor[HTML]{C6EFCE}{0.297} & 0.302 & \cellcolor[HTML]{C6EFCE}{0.194} & 0.146 & \cellcolor[HTML]{C6EFCE}{0.043} & 0.231 & \cellcolor[HTML]{C6EFCE}{0.068} & 16.882 & \cellcolor[HTML]{C6EFCE}{4.879} & 27.133 & \cellcolor[HTML]{C6EFCE}{8.710}\\
& 30 & 4.460 & \cellcolor[HTML]{C6EFCE}{2.552} & 19.101 & \cellcolor[HTML]{C6EFCE}{2.742} & 0.325 & \cellcolor[HTML]{C6EFCE}{0.312} & 0.237 & 0.250 & 0.057 & \cellcolor[HTML]{C6EFCE}{0.047} & 0.178 & \cellcolor[HTML]{C6EFCE}{0.064} & 16.882 & \cellcolor[HTML]{C6EFCE}{5.411} & 27.566 & \cellcolor[HTML]{C6EFCE}{4.873}\\
& 40 & 4.338 & \cellcolor[HTML]{C6EFCE}{2.095} & 6.447 & \cellcolor[HTML]{C6EFCE}{2.355} & 0.325 & 0.326 & 0.244 & \cellcolor[HTML]{C6EFCE}{0.197} & 0.160 & \cellcolor[HTML]{C6EFCE}{0.047} & 0.132 & \cellcolor[HTML]{C6EFCE}{0.070} & 16.882 & \cellcolor[HTML]{C6EFCE}{4.761} & 28.756 & \cellcolor[HTML]{C6EFCE}{11.436}\\
& 50 & 4.419 & \cellcolor[HTML]{C6EFCE}{2.361} & 7.511 & \cellcolor[HTML]{C6EFCE}{2.343} & 0.325 & \cellcolor[HTML]{C6EFCE}{0.290} & 0.236 & 0.241 & 0.160 & \cellcolor[HTML]{C6EFCE}{0.050} & 0.140 & \cellcolor[HTML]{C6EFCE}{0.052} & 16.882 & \cellcolor[HTML]{C6EFCE}{3.711} & 30.036 & \cellcolor[HTML]{C6EFCE}{3.770}\\
& 60 & 4.402 & \cellcolor[HTML]{C6EFCE}{3.773} & 5.628 & \cellcolor[HTML]{C6EFCE}{2.703} & 0.325 & \cellcolor[HTML]{C6EFCE}{0.278} & 0.237 & \cellcolor[HTML]{C6EFCE}{0.192} & 0.180 & \cellcolor[HTML]{C6EFCE}{0.045} & 0.131 & \cellcolor[HTML]{C6EFCE}{0.054} & 16.882 & \cellcolor[HTML]{C6EFCE}{4.316} & 29.280 & \cellcolor[HTML]{C6EFCE}{3.173}\\
& 70 & 4.412 & \cellcolor[HTML]{C6EFCE}{2.431} & 4.996 & \cellcolor[HTML]{C6EFCE}{2.329} & 0.325 & \cellcolor[HTML]{C6EFCE}{0.288} & 0.302 & \cellcolor[HTML]{C6EFCE}{0.231} & 0.152 & \cellcolor[HTML]{C6EFCE}{0.048} & 0.097 & \cellcolor[HTML]{C6EFCE}{0.048} & 16.882 & \cellcolor[HTML]{C6EFCE}{4.061} & 30.322 & \cellcolor[HTML]{C6EFCE}{3.164}\\
\midrule
\multirow{7}{*}{Flash}
& 10 & 4.414 & 4.620 & 7.171 & \cellcolor[HTML]{C6EFCE}{4.790} & 0.401 & \cellcolor[HTML]{C6EFCE}{0.313} & 0.251 & \cellcolor[HTML]{C6EFCE}{0.196} & 0.169 & \cellcolor[HTML]{C6EFCE}{0.057} & 0.316 & \cellcolor[HTML]{C6EFCE}{0.132} & 39.572 & \cellcolor[HTML]{C6EFCE}{18.528} & 33.141 & \cellcolor[HTML]{C6EFCE}{21.917}\\
& 20 & 6.446 & \cellcolor[HTML]{C6EFCE}{2.450} & 6.052 & \cellcolor[HTML]{C6EFCE}{3.097} & 0.356 & \cellcolor[HTML]{C6EFCE}{0.301} & 0.350 & \cellcolor[HTML]{C6EFCE}{0.252} & 0.153 & \cellcolor[HTML]{C6EFCE}{0.152} & 0.127 & \cellcolor[HTML]{C6EFCE}{0.116} & 39.002 & \cellcolor[HTML]{C6EFCE}{20.731} & 36.251 & \cellcolor[HTML]{C6EFCE}{20.082}\\
& 30 & 4.437 & \cellcolor[HTML]{C6EFCE}{1.887} & 8.227 & \cellcolor[HTML]{C6EFCE}{2.718} & 0.368 & \cellcolor[HTML]{C6EFCE}{0.302} & 0.256 & \cellcolor[HTML]{C6EFCE}{0.214} & 0.140 & \cellcolor[HTML]{C6EFCE}{0.052} & 0.140 & \cellcolor[HTML]{C6EFCE}{0.064} & 39.275 & \cellcolor[HTML]{C6EFCE}{19.485} & 22.261 & \cellcolor[HTML]{C6EFCE}{17.760}\\
& 40 & 4.478 & \cellcolor[HTML]{C6EFCE}{1.865} & 6.017 & \cellcolor[HTML]{C6EFCE}{2.166} & 0.330 & \cellcolor[HTML]{C6EFCE}{0.303} & 0.252 & \cellcolor[HTML]{C6EFCE}{0.191} & 0.151 & \cellcolor[HTML]{C6EFCE}{0.054} & 0.135 & \cellcolor[HTML]{C6EFCE}{0.074} & 31.472 & \cellcolor[HTML]{C6EFCE}{18.439} & 15.941 & 21.517\\
& 50 & 4.567 & \cellcolor[HTML]{C6EFCE}{1.848} & 5.480 & \cellcolor[HTML]{C6EFCE}{1.743} & 0.379 & \cellcolor[HTML]{C6EFCE}{0.311} & 0.385 & \cellcolor[HTML]{C6EFCE}{0.195} & 0.155 & \cellcolor[HTML]{C6EFCE}{0.073} & 0.146 & 0.148 & 31.053 & \cellcolor[HTML]{C6EFCE}{19.869} & 34.279 & \cellcolor[HTML]{C6EFCE}{19.243}\\
& 60 & 4.891 & \cellcolor[HTML]{C6EFCE}{1.842} & 4.816 & \cellcolor[HTML]{C6EFCE}{1.885} & 0.363 & \cellcolor[HTML]{C6EFCE}{0.341} & 0.359 & \cellcolor[HTML]{C6EFCE}{0.224} & 0.143 & \cellcolor[HTML]{C6EFCE}{0.053} & 0.165 & \cellcolor[HTML]{C6EFCE}{0.066} & 31.663 & \cellcolor[HTML]{C6EFCE}{19.213} & 16.717 & 19.693\\
& 70 & 4.956 & \cellcolor[HTML]{C6EFCE}{1.762} & 5.519 & \cellcolor[HTML]{C6EFCE}{1.927} & 0.373 & \cellcolor[HTML]{C6EFCE}{0.294} & 0.314 & \cellcolor[HTML]{C6EFCE}{0.207} & 0.153 & \cellcolor[HTML]{C6EFCE}{0.145} & 0.152 & \cellcolor[HTML]{C6EFCE}{0.097} & 18.694 & \cellcolor[HTML]{C6EFCE}{10.043} & 18.398 & \cellcolor[HTML]{C6EFCE}{14.607}\\
\midrule
\multirow{7}{*}{CoMSA}
& 10 & 4.264 & 4.491 & 7.238 & \cellcolor[HTML]{C6EFCE}{3.742} & 0.647 & \cellcolor[HTML]{C6EFCE}{0.336} & 0.423 & \cellcolor[HTML]{C6EFCE}{0.218} & 0.178 & \cellcolor[HTML]{C6EFCE}{0.159} & 0.126 & \cellcolor[HTML]{C6EFCE}{0.101} & 17.284 & \cellcolor[HTML]{C6EFCE}{8.363} & 31.829 & \cellcolor[HTML]{C6EFCE}{20.727}\\
& 20 & 2.712 & 3.216 & 6.303 & \cellcolor[HTML]{C6EFCE}{2.995} & 0.506 & \cellcolor[HTML]{C6EFCE}{0.300} & 0.373 & \cellcolor[HTML]{C6EFCE}{0.206} & 0.049 & 0.049 & 0.202 & \cellcolor[HTML]{C6EFCE}{0.049} & 12.847 & \cellcolor[HTML]{C6EFCE}{6.074} & 11.864 & \cellcolor[HTML]{C6EFCE}{4.900}\\
& 30 & 2.817 & \cellcolor[HTML]{C6EFCE}{2.712} & 5.766 & \cellcolor[HTML]{C6EFCE}{2.608} & 0.536 & \cellcolor[HTML]{C6EFCE}{0.323} & 0.495 & \cellcolor[HTML]{C6EFCE}{0.203} & 0.055 & \cellcolor[HTML]{C6EFCE}{0.052} & 0.046 & \cellcolor[HTML]{C6EFCE}{0.044} & 6.932 & \cellcolor[HTML]{C6EFCE}{4.700} & 3.736 & 3.940\\
& 40 & 2.547 & 2.796 & 4.265 & \cellcolor[HTML]{C6EFCE}{2.544} & 0.472 & \cellcolor[HTML]{C6EFCE}{0.318} & 0.348 & \cellcolor[HTML]{C6EFCE}{0.205} & 0.044 & 0.046 & 0.055 & \cellcolor[HTML]{C6EFCE}{0.040} & 5.752 & \cellcolor[HTML]{C6EFCE}{4.074} & 5.414 & \cellcolor[HTML]{C6EFCE}{3.994}\\
& 50 & 2.430 & \cellcolor[HTML]{C6EFCE}{1.936} & 3.931 & \cellcolor[HTML]{C6EFCE}{1.963} & 0.576 & \cellcolor[HTML]{C6EFCE}{0.271} & 0.468 & \cellcolor[HTML]{C6EFCE}{0.204} & 0.046 & \cellcolor[HTML]{C6EFCE}{0.045} & 0.113 & \cellcolor[HTML]{C6EFCE}{0.040} & 5.922 & \cellcolor[HTML]{C6EFCE}{3.657} & 4.613 & 4.618\\
& 60 & 2.097 & \cellcolor[HTML]{C6EFCE}{1.797} & 2.778 & \cellcolor[HTML]{C6EFCE}{1.801} & 0.562 & \cellcolor[HTML]{C6EFCE}{0.268} & 0.434 & \cellcolor[HTML]{C6EFCE}{0.227} & 0.048 & \cellcolor[HTML]{C6EFCE}{0.043} & 0.115 & \cellcolor[HTML]{C6EFCE}{0.065} & 4.934 & \cellcolor[HTML]{C6EFCE}{3.611} & 3.863 & \cellcolor[HTML]{C6EFCE}{3.043}\\
& 70 & 1.931 & \cellcolor[HTML]{C6EFCE}{1.744} & 3.073 & \cellcolor[HTML]{C6EFCE}{1.845} & 0.548 & \cellcolor[HTML]{C6EFCE}{0.265} & 0.355 & \cellcolor[HTML]{C6EFCE}{0.184} & 0.044 & \cellcolor[HTML]{C6EFCE}{0.042} & 0.038 & 0.041 & 4.934 & \cellcolor[HTML]{C6EFCE}{3.522} & 3.631 & \cellcolor[HTML]{C6EFCE}{2.554}\\
\midrule
\multirow{7}{*}{NSBS}
& 10 & 2.581 & 3.817 & 8.686 & \cellcolor[HTML]{C6EFCE}{4.416} & 0.586 & \cellcolor[HTML]{C6EFCE}{0.527} & 0.964 & \cellcolor[HTML]{C6EFCE}{0.352} & 0.523 & \cellcolor[HTML]{C6EFCE}{0.515} & 0.288 & \cellcolor[HTML]{C6EFCE}{0.254} & 9.106 & \cellcolor[HTML]{C6EFCE}{8.939} & 28.399 & \cellcolor[HTML]{C6EFCE}{23.359}\\
& 20 & 2.380 & \cellcolor[HTML]{C6EFCE}{2.369} & 3.444 & \cellcolor[HTML]{C6EFCE}{2.708} & 0.341 & \cellcolor[HTML]{C6EFCE}{0.335} & 0.282 & \cellcolor[HTML]{C6EFCE}{0.270} & 0.056 & \cellcolor[HTML]{C6EFCE}{0.053} & 0.064 & 0.067 & 7.384 & \cellcolor[HTML]{C6EFCE}{5.879} & 5.101 & \cellcolor[HTML]{C6EFCE}{3.569}\\
& 30 & 2.460 & \cellcolor[HTML]{C6EFCE}{2.284} & 5.170 & \cellcolor[HTML]{C6EFCE}{2.458} & 0.312 & \cellcolor[HTML]{C6EFCE}{0.273} & 0.251 & \cellcolor[HTML]{C6EFCE}{0.214} & 0.052 & \cellcolor[HTML]{C6EFCE}{0.046} & 0.041 & \cellcolor[HTML]{C6EFCE}{0.037} & 6.119 & \cellcolor[HTML]{C6EFCE}{4.308} & 4.278 & \cellcolor[HTML]{C6EFCE}{3.507}\\
& 40 & 2.549 & \cellcolor[HTML]{C6EFCE}{1.730} & 3.124 & \cellcolor[HTML]{C6EFCE}{2.188} & 0.321 & \cellcolor[HTML]{C6EFCE}{0.285} & 0.269 & \cellcolor[HTML]{C6EFCE}{0.221} & 0.047 & \cellcolor[HTML]{C6EFCE}{0.045} & 0.034 & \cellcolor[HTML]{C6EFCE}{0.032} & 4.393 & \cellcolor[HTML]{C6EFCE}{3.638} & 4.072 & \cellcolor[HTML]{C6EFCE}{2.906}\\
& 50 & 2.097 & \cellcolor[HTML]{C6EFCE}{1.892} & 2.785 & \cellcolor[HTML]{C6EFCE}{2.217} & 0.328 & \cellcolor[HTML]{C6EFCE}{0.298} & 0.226 & \cellcolor[HTML]{C6EFCE}{0.196} & 0.044 & \cellcolor[HTML]{C6EFCE}{0.042} & 0.034 & \cellcolor[HTML]{C6EFCE}{0.031} & 4.084 & \cellcolor[HTML]{C6EFCE}{3.336} & 4.574 & \cellcolor[HTML]{C6EFCE}{2.581}\\
& 60 & 1.874 & 1.920 & 2.922 & \cellcolor[HTML]{C6EFCE}{2.178} & 0.391 & \cellcolor[HTML]{C6EFCE}{0.267} & 0.271 & \cellcolor[HTML]{C6EFCE}{0.190} & 0.044 & \cellcolor[HTML]{C6EFCE}{0.043} & 0.034 & \cellcolor[HTML]{C6EFCE}{0.030} & 4.183 & \cellcolor[HTML]{C6EFCE}{3.321} & 2.679 & \cellcolor[HTML]{C6EFCE}{2.650}\\
& 70 & 1.765 & 1.784 & 2.065 & \cellcolor[HTML]{C6EFCE}{1.993} & 0.298 & \cellcolor[HTML]{C6EFCE}{0.295} & 0.218 & \cellcolor[HTML]{C6EFCE}{0.199} & 0.042 & \cellcolor[HTML]{C6EFCE}{0.040} & 0.032 & 0.033 & 4.048 & \cellcolor[HTML]{C6EFCE}{3.114} & 2.936 & \cellcolor[HTML]{C6EFCE}{2.151}\\
\midrule
\multirow{7}{*}{EHVI}
& 10 & 6.982 & \cellcolor[HTML]{C6EFCE}{3.573} & 7.654 & \cellcolor[HTML]{C6EFCE}{2.960} & 0.474 & \cellcolor[HTML]{C6EFCE}{0.278} & 0.237 & \cellcolor[HTML]{C6EFCE}{0.204} & 0.148 & \cellcolor[HTML]{C6EFCE}{0.056} & 0.222 & \cellcolor[HTML]{C6EFCE}{0.073} & 45.765 & \cellcolor[HTML]{C6EFCE}{23.719} & 38.815 & \cellcolor[HTML]{C6EFCE}{24.444}\\
& 20 & 6.111 & \cellcolor[HTML]{C6EFCE}{3.579} & 6.588 & \cellcolor[HTML]{C6EFCE}{2.501} & 0.369 & \cellcolor[HTML]{C6EFCE}{0.286} & 0.212 & \cellcolor[HTML]{C6EFCE}{0.206} & 0.059 & \cellcolor[HTML]{C6EFCE}{0.053} & 0.075 & 0.095 & 39.563 & \cellcolor[HTML]{C6EFCE}{19.075} & 20.084 & \cellcolor[HTML]{C6EFCE}{5.454}\\
& 30 & 4.835 & \cellcolor[HTML]{C6EFCE}{3.399} & 6.280 & \cellcolor[HTML]{C6EFCE}{2.679} & 0.356 & \cellcolor[HTML]{C6EFCE}{0.314} & 0.231 & \cellcolor[HTML]{C6EFCE}{0.201} & 0.056 & \cellcolor[HTML]{C6EFCE}{0.055} & 0.114 & \cellcolor[HTML]{C6EFCE}{0.050} & 26.035 & \cellcolor[HTML]{C6EFCE}{6.973} & 12.896 & \cellcolor[HTML]{C6EFCE}{7.255}\\
& 40 & 2.405 & \cellcolor[HTML]{C6EFCE}{2.306} & 4.405 & \cellcolor[HTML]{C6EFCE}{2.357} & 0.369 & \cellcolor[HTML]{C6EFCE}{0.271} & 0.249 & \cellcolor[HTML]{C6EFCE}{0.198} & 0.056 & \cellcolor[HTML]{C6EFCE}{0.054} & 0.065 & \cellcolor[HTML]{C6EFCE}{0.058} & 23.990 & \cellcolor[HTML]{C6EFCE}{7.239} & 11.428 & \cellcolor[HTML]{C6EFCE}{4.485}\\
& 50 & 2.418 & \cellcolor[HTML]{C6EFCE}{1.777} & 2.755 & \cellcolor[HTML]{C6EFCE}{2.130} & 0.387 & \cellcolor[HTML]{C6EFCE}{0.270} & 0.247 & \cellcolor[HTML]{C6EFCE}{0.202} & 0.052 & \cellcolor[HTML]{C6EFCE}{0.049} & 0.113 & \cellcolor[HTML]{C6EFCE}{0.040} & 23.046 & \cellcolor[HTML]{C6EFCE}{6.224} & 7.166 & \cellcolor[HTML]{C6EFCE}{3.674}\\
& 60 & 2.316 & \cellcolor[HTML]{C6EFCE}{1.793} & 2.611 & \cellcolor[HTML]{C6EFCE}{2.261} & 0.331 & \cellcolor[HTML]{C6EFCE}{0.283} & 0.213 & 0.215 & 0.054 & \cellcolor[HTML]{C6EFCE}{0.046} & 0.055 & \cellcolor[HTML]{C6EFCE}{0.044} & 23.093 & \cellcolor[HTML]{C6EFCE}{4.652} & 18.599 & \cellcolor[HTML]{C6EFCE}{3.702}\\
& 70 & 2.192 & \cellcolor[HTML]{C6EFCE}{1.794} & 2.550 & \cellcolor[HTML]{C6EFCE}{1.916} & 0.329 & \cellcolor[HTML]{C6EFCE}{0.284} & 0.230 & \cellcolor[HTML]{C6EFCE}{0.200} & 0.049 & \cellcolor[HTML]{C6EFCE}{0.044} & 0.118 & \cellcolor[HTML]{C6EFCE}{0.041} & 23.363 & \cellcolor[HTML]{C6EFCE}{5.070} & 9.497 & \cellcolor[HTML]{C6EFCE}{3.865}\\
\midrule
\multirow{7}{*}{TSEMO}
& 10 & 5.487 & \cellcolor[HTML]{C6EFCE}{4.013} & 7.393 & \cellcolor[HTML]{C6EFCE}{3.529} & 0.370 & \cellcolor[HTML]{C6EFCE}{0.281} & 0.233 & \cellcolor[HTML]{C6EFCE}{0.201} & 0.057 & \cellcolor[HTML]{C6EFCE}{0.050} & 0.314 & \cellcolor[HTML]{C6EFCE}{0.113} & 22.678 & 23.882 & 33.315 & \cellcolor[HTML]{C6EFCE}{26.562}\\
& 20 & 2.735 & \cellcolor[HTML]{C6EFCE}{2.464} & 6.698 & \cellcolor[HTML]{C6EFCE}{2.997} & 0.358 & \cellcolor[HTML]{C6EFCE}{0.300} & 0.207 & \cellcolor[HTML]{C6EFCE}{0.199} & 0.048 & 0.049 & 0.064 & \cellcolor[HTML]{C6EFCE}{0.044} & 22.684 & \cellcolor[HTML]{C6EFCE}{8.223} & 23.926 & \cellcolor[HTML]{C6EFCE}{10.983}\\
& 30 & 5.328 & \cellcolor[HTML]{C6EFCE}{2.370} & 4.411 & \cellcolor[HTML]{C6EFCE}{2.683} & 0.357 & \cellcolor[HTML]{C6EFCE}{0.293} & 0.249 & \cellcolor[HTML]{C6EFCE}{0.199} & 0.047 & \cellcolor[HTML]{C6EFCE}{0.045} & 0.053 & \cellcolor[HTML]{C6EFCE}{0.040} & 23.268 & \cellcolor[HTML]{C6EFCE}{12.335} & 20.878 & \cellcolor[HTML]{C6EFCE}{6.156}\\
& 40 & 2.693 & \cellcolor[HTML]{C6EFCE}{1.828} & 3.873 & \cellcolor[HTML]{C6EFCE}{2.232} & 0.337 & \cellcolor[HTML]{C6EFCE}{0.286} & 0.267 & \cellcolor[HTML]{C6EFCE}{0.196} & 0.048 & \cellcolor[HTML]{C6EFCE}{0.044} & 0.042 & \cellcolor[HTML]{C6EFCE}{0.032} & 22.469 & \cellcolor[HTML]{C6EFCE}{17.618} & 8.006 & \cellcolor[HTML]{C6EFCE}{6.819}\\
& 50 & 1.982 & \cellcolor[HTML]{C6EFCE}{1.919} & 3.530 & \cellcolor[HTML]{C6EFCE}{2.137} & 0.329 & \cellcolor[HTML]{C6EFCE}{0.274} & 0.211 & \cellcolor[HTML]{C6EFCE}{0.193} & 0.046 & \cellcolor[HTML]{C6EFCE}{0.046} & 0.037 & \cellcolor[HTML]{C6EFCE}{0.034} & 22.485 & \cellcolor[HTML]{C6EFCE}{8.047} & 16.673 & \cellcolor[HTML]{C6EFCE}{6.320}\\
& 60 & 1.776 & 1.874 & 3.304 & \cellcolor[HTML]{C6EFCE}{2.126} & 0.330 & \cellcolor[HTML]{C6EFCE}{0.279} & 0.202 & \cellcolor[HTML]{C6EFCE}{0.192} & 0.046 & \cellcolor[HTML]{C6EFCE}{0.043} & 0.032 & 0.033 & 23.295 & \cellcolor[HTML]{C6EFCE}{8.256} & 6.237 & \cellcolor[HTML]{C6EFCE}{5.102}\\
& 70 & 1.722 & 1.816 & 2.544 & \cellcolor[HTML]{C6EFCE}{2.213} & 0.333 & \cellcolor[HTML]{C6EFCE}{0.290} & 0.233 & \cellcolor[HTML]{C6EFCE}{0.191} & 0.045 & \cellcolor[HTML]{C6EFCE}{0.040} & 0.044 & \cellcolor[HTML]{C6EFCE}{0.036} & 23.300 & \cellcolor[HTML]{C6EFCE}{7.829} & 12.383 & \cellcolor[HTML]{C6EFCE}{5.115}\\
\midrule
\multirow{7}{*}{NSGA-III}
& 10 & 5.672 & \cellcolor[HTML]{C6EFCE}{4.459} & 6.537 & \cellcolor[HTML]{C6EFCE}{3.876} & 0.447 & \cellcolor[HTML]{C6EFCE}{0.281} & 0.441 & \cellcolor[HTML]{C6EFCE}{0.238} & 0.160 & \cellcolor[HTML]{C6EFCE}{0.052} & 0.105 & \cellcolor[HTML]{C6EFCE}{0.102} & 41.769 & \cellcolor[HTML]{C6EFCE}{27.260} & 33.538 & \cellcolor[HTML]{C6EFCE}{15.137}\\
& 20 & 3.364 & \cellcolor[HTML]{C6EFCE}{3.024} & 5.045 & \cellcolor[HTML]{C6EFCE}{2.974} & 0.512 & \cellcolor[HTML]{C6EFCE}{0.296} & 0.306 & \cellcolor[HTML]{C6EFCE}{0.211} & 0.153 & \cellcolor[HTML]{C6EFCE}{0.052} & 0.112 & \cellcolor[HTML]{C6EFCE}{0.084} & 39.967 & \cellcolor[HTML]{C6EFCE}{7.540} & 24.721 & \cellcolor[HTML]{C6EFCE}{4.517}\\
& 30 & 2.611 & \cellcolor[HTML]{C6EFCE}{2.450} & 6.489 & \cellcolor[HTML]{C6EFCE}{2.463} & 0.385 & \cellcolor[HTML]{C6EFCE}{0.300} & 0.430 & \cellcolor[HTML]{C6EFCE}{0.226} & 0.154 & \cellcolor[HTML]{C6EFCE}{0.046} & 0.091 & \cellcolor[HTML]{C6EFCE}{0.047} & 27.023 & \cellcolor[HTML]{C6EFCE}{7.103} & 5.093 & \cellcolor[HTML]{C6EFCE}{3.179}\\
& 40 & 2.367 & \cellcolor[HTML]{C6EFCE}{2.101} & 2.988 & \cellcolor[HTML]{C6EFCE}{2.192} & 0.351 & \cellcolor[HTML]{C6EFCE}{0.314} & 0.392 & \cellcolor[HTML]{C6EFCE}{0.204} & 0.054 & \cellcolor[HTML]{C6EFCE}{0.045} & 0.035 & \cellcolor[HTML]{C6EFCE}{0.033} & 39.058 & \cellcolor[HTML]{C6EFCE}{6.423} & 18.390 & \cellcolor[HTML]{C6EFCE}{3.045}\\
& 50 & 2.389 & \cellcolor[HTML]{C6EFCE}{2.349} & 3.668 & \cellcolor[HTML]{C6EFCE}{2.022} & 0.293 & 0.307 & 0.215 & \cellcolor[HTML]{C6EFCE}{0.204} & 0.056 & \cellcolor[HTML]{C6EFCE}{0.045} & 0.075 & \cellcolor[HTML]{C6EFCE}{0.042} & 4.440 & 5.203 & 6.035 & \cellcolor[HTML]{C6EFCE}{2.869}\\
& 60 & 2.710 & \cellcolor[HTML]{C6EFCE}{2.028} & 3.081 & \cellcolor[HTML]{C6EFCE}{2.482} & 0.345 & \cellcolor[HTML]{C6EFCE}{0.289} & 0.241 & \cellcolor[HTML]{C6EFCE}{0.197} & 0.052 & \cellcolor[HTML]{C6EFCE}{0.048} & 0.043 & \cellcolor[HTML]{C6EFCE}{0.031} & 7.959 & \cellcolor[HTML]{C6EFCE}{6.804} & 5.097 & \cellcolor[HTML]{C6EFCE}{3.389}\\
& 70 & 2.317 & \cellcolor[HTML]{C6EFCE}{1.846} & 3.062 & \cellcolor[HTML]{C6EFCE}{2.127} & 0.302 & \cellcolor[HTML]{C6EFCE}{0.289} & 0.228 & \cellcolor[HTML]{C6EFCE}{0.199} & 0.052 & \cellcolor[HTML]{C6EFCE}{0.045} & 0.055 & \cellcolor[HTML]{C6EFCE}{0.049} & 7.429 & \cellcolor[HTML]{C6EFCE}{3.790} & 4.663 & \cellcolor[HTML]{C6EFCE}{2.922}\\

\bottomrule
\end{tabular}
}
\end{table*}

\noindent \textbf{\textit{Results.}}
\textbf{\tool space pruning capability enhances the effectiveness of baselines.}
The results in Table~\ref{tab:filtering_effectiveness} show that the space pruning provides a consistent benefit, improving the effectiveness of the baselines. To quantify this, we conducted \sout{336} \rv{448} evaluation scenarios (4 systems $\times$ \sout{6} \rv{8} baselines $\times$ 7 sample sizes $\times$ 2 models). As shown in the table, the models trained on the pruned space achieved a lower (better) RMSE in \sout{295} \rv{410} of these scenarios, an improvement rate of \sout{87\%} \rv{91.5\%}. The performance gain can be dramatic; for example, on the \texttt{SQLite} benchmark (XGBoost, size 10), pruning allows \texttt{Random} to reduce its RMSE by nearly 87\% (from 0.515 to 0.067). This confirms that \tool's space pruning is a generally effective technique that creates a more focused search space for any subsequent sampling method.

\rv{\textbf{\tool's adaptive strategy refinement dynamically focuses on informative regions.}} \rv{The effectiveness of \tool is not just due to static pruning, but also its ability to dynamically learn from performance data. As illustrated by the \texttt{LRZIP} \textbf{Running Example} (Section ~\ref{sec:methodology}), the framework's feedback loop enables this. Initially, the \textbf{Sampling Strategy Designer} uses a broad, coverage-based strategy. However, after the first iteration, the \textbf{Performance Trend Analyzer} identifies critical insights from the feedback, such as \texttt{algorithm -g} being a performance anomaly and \texttt{Compression level (-L)} having low sensitivity.}
\rv{This analysis immediately triggers a strategic shift. The \textbf{Sampling Strategy Designer} refines the plan to focus sampling on the \texttt{-g} anomaly while simultaneously de-prioritizing (or ``pruning'') the low-sensitivity \texttt{-L} option to improve efficiency. This step-by-step trace demonstrates that \tool does not rely on a fixed heuristic. Instead, it uses its reasoning capabilities to interpret feedback, identify high-impact parameters and anomalies, and adaptively concentrate the search on the most informative regions of the configuration space, which is key to building better performance models.}

\greybox{\textbf{RQ2 Summary:} Our analysis shows \tool's effectiveness stems from two mechanisms. First, its static pruning of the configuration space is a generally effective technique, improving baseline performance in \textbf{91.5\%} of scenarios. Second, its dynamic feedback loop allows it to learn from observed data to identify anomalies and adaptively focus sampling on these performance-critical regions.}

\subsection*{RQ3: How does LLMs selection impact \tool's effectiveness?}
\label{sec:rq3}

\noindent \textbf{\textit{Motivation.}}
\tool's sampling process involves three core components: the \textbf{Strategy Designer}, the \textbf{Performance Analyzer}, and the \textbf{Configuration Generator}. The \textbf{Designer} and \textbf{Analyzer} require reasoning to interpret performance feedback and design strategies. In contrast, the \textbf{Generator} performs the more straightforward task of translating a sampling strategy into configurations. This separation allows for a hybrid model approach. Understanding how the LLM choices for each component within \tool affect its overall effectiveness is crucial for optimizing the \tool's cost-effectiveness.
Therefore, in this section, we evaluate the impact of LLM choices from two perspectives. First, we assess the performance of LLMs with different reasoning capabilities when serving \textbf{Designer} and \textbf{Analyzer}. Second, we analyze the correlation between model size and effectiveness to investigate whether larger models are more effective for the \textbf{Generator}.

\noindent \textbf{\textit{Approach.}} Our approach for this analysis consists of two distinct experiments. First, to evaluate the high-reasoning Designer and Analyzer, we compare the performance of four models: \texttt{GPT-4o}, \texttt{Deepseek-R1}, \texttt{Llama 3.3} and \texttt{Qwen2.5-72B}. \sout{Second, to analyze the relationship between model size and effectiveness for the Generator, we compare six different models, including several of varying sizes from the same family (e.g., the \texttt{Qwen2.5} series).} \rv{Second, to analyze the relationship between effectiveness, model size, and architecture for the Generator, we compare a diverse set of nine models. This selection was made to evaluate two key factors: 1) the impact of model scale, by including varying sizes from the same families (e.g., the \texttt{Qwen2.5}, \texttt{Llama3.1}, and \texttt{DeepSeek-R1} series), and 2) the impact of architecture, by comparing traditional dense models (the \texttt{Qwen2.5} and \texttt{Llama3.1} families) against Mixture of Experts (MoE) models (the \texttt{DeepSeek-R1} and \texttt{Mixtral} families).} For both experiments, each model is tested across sample sizes ranging from 10 to 70, and each run is repeated ten times to ensure statistical robustness.

% \begin{table}[htbp]
% \centering
% \caption{(RQ3) The Impact of LLM Choice of Reasoning Roles on Sampling Effectiveness for LRZIP Compression Time (RMSE)}
% \label{tab:different_model}
% \resizebox{\columnwidth}{!}{
%     \begin{tabular}{l | r r r r | r r r r}
%     \toprule
%     \multirow{3}{*}{\makecell{Sample \\ Size}} & \multicolumn{8}{c}{Compression Time} \\
%     \cmidrule(lr){2-9}
%     & \multicolumn{4}{c|}{XGBoost} & \multicolumn{4}{c}{DeepPerf} \\
%     \cmidrule(lr){2-5} \cmidrule(lr){6-9}
%     & GPT-4o & Deepseek-R1 & Llama 3.3 & Qwen2.5-72B & GPT-4o & Deepseek-R1 & Llama 3.3 & Qwen2.5-72B \\
%     \midrule
%     10 & 2.486 & \cellcolor[HTML]{C6EFCE}{2.189} & 2.245 & 2.935 & 2.328 & \cellcolor[HTML]{C6EFCE}{2.318} & 3.528 & 2.488 \\
%     20 & 2.392 & 1.867 & \cellcolor[HTML]{C6EFCE}{1.729} & 2.387 & 2.642 & 2.449 & 2.641 & \cellcolor[HTML]{C6EFCE}{2.350} \\
%     30 & 2.200 & \cellcolor[HTML]{C6EFCE}{1.660} & 1.755 & 1.853 & 2.158 & \cellcolor[HTML]{C6EFCE}{1.783} & 2.044 & 2.125 \\
%     40 & 2.037 & \cellcolor[HTML]{C6EFCE}{1.684} & 1.905 & 1.928 & \cellcolor[HTML]{C6EFCE}{1.898} & 1.899 & 2.067 & 2.074 \\
%     50 & \cellcolor[HTML]{C6EFCE}{1.683} & 1.870 & 1.789 & 1.969 & 1.875 & 2.133 & 1.812 & \cellcolor[HTML]{C6EFCE}{1.738} \\
%     60 & 1.821 & 1.832 & \cellcolor[HTML]{C6EFCE}{1.755} & 1.809 & 1.995 & \cellcolor[HTML]{C6EFCE}{1.826} & 2.006 & 1.877 \\
%     70 & 1.785 & \cellcolor[HTML]{C6EFCE}{1.728} & 1.795 & 1.798 & 1.810 & \cellcolor[HTML]{C6EFCE}{1.529} & 1.913 & 1.892 \\

%     \bottomrule
%     \end{tabular}
% }
% \end{table}

\begin{table}[htbp]
\centering
\caption{(RQ3) Impact of LLM Choice on Sampling Effectiveness for LRZIP Compression Time on XGBoost Model}
\label{tab:different_model}
\resizebox{0.8\columnwidth}{!}{
\renewcommand{\arraystretch}{0.5}
\begin{tabular}{l | r r r r}
\toprule
\multirow{2}{*}{\makecell{Sample \\ Size}} & \multicolumn{4}{c}{Compression Time (XGBoost)} \\
\cmidrule(lr){2-5}
& GPT-4o & Deepseek-R1 & Llama 3.3 & Qwen2.5-72B \\
\midrule
10 & 2.486 & \cellcolor[HTML]{C6EFCE}{2.189} & 2.245 & 2.935 \\
20 & 2.392 & 1.867 & \cellcolor[HTML]{C6EFCE}{1.729} & 2.387 \\
30 & 2.200 & \cellcolor[HTML]{C6EFCE}{1.660} & 1.755 & 1.853 \\
40 & 2.037 & \cellcolor[HTML]{C6EFCE}{1.684} & 1.905 & 1.928 \\
50 & \cellcolor[HTML]{C6EFCE}{1.683} & 1.870 & 1.789 & 1.969 \\
60 & 1.821 & 1.832 & \cellcolor[HTML]{C6EFCE}{1.755} & 1.809 \\
70 & 1.785 & \cellcolor[HTML]{C6EFCE}{1.728} & 1.795 & 1.798 \\
\bottomrule
\end{tabular}
}
\end{table}

\begin{table*}[htbp]
\centering
\caption{Impact of LLM Choice on Sampling Effectiveness for LRZIP Compression Time (RMSE, XGBoost)}
\label{tab:different_gen_model}

\resizebox{\textwidth}{!}{%
\setlength{\tabcolsep}{12pt}
\renewcommand{\arraystretch}{0.6}
\begin{tabular}{l | c c c c c c c c c}
\toprule
\multirow{2}{*}{\makecell{Sample \\ Size}} & \multicolumn{9}{c}{Compression Time (XGBoost)} \\
\cmidrule(lr){2-10}
& Qwen2.5-72B & Qwen2.5-32B & Qwen2.5-7B & Llama3.1-70B & Llama3.1-8B & DeepSeek-R1-70B & DeepSeek-R1-32B & Mixtral-8x22B & Mixtral-8x7B \\
\midrule
10 & 2.309 & 2.829 & 2.552 & \cellcolor[HTML]{C6EFCE}{2.189} & 3.190 & 2.334 & 2.419 & 2.350 & 2.643 \\
20 & 2.177 & 2.361 & 2.484 & 1.959 & 1.984 & 1.840 & 2.406 & \cellcolor[HTML]{C6EFCE}{1.818} & 2.090 \\
30 & 1.877 & \cellcolor[HTML]{C6EFCE}{1.743} & 1.831 & 2.062 & 2.220 & 1.838 & 2.041 & 2.156 & 2.352 \\
40 & 1.765 & \cellcolor[HTML]{C6EFCE}{1.738} & 1.828 & 1.903 & 2.168 & 1.936 & 1.787 & 1.873 & 1.757 \\
50 & \cellcolor[HTML]{C6EFCE}{1.723} & 1.867 & 2.171 & 1.773 & 1.852 & 1.936 & 1.932 & 1.725 & 1.947 \\
60 & 1.784 & 1.759 & 2.040 & \cellcolor[HTML]{C6EFCE}{1.739} & 1.938 & 1.896 & 1.742 & 1.836 & 1.877 \\
70 & 1.763 & 1.756 & 1.745 & 1.799 & 1.913 & 1.928 & 1.958 & 1.748 & \cellcolor[HTML]{C6EFCE}{1.697} \\
\bottomrule
\end{tabular}
}
\end{table*}

\noindent \textbf{\textit{Results.}}
\textbf{Within \tool, the \textbf{Strategy Designer} and \textbf{Performance Analyzer} leverage LLMs with stronger reasoning capabilities can deliver superior performance.}
% Our results indicate that the effectiveness of \tool is closely tied to the advanced reasoning capabilities of the used LLMs. 
Models with stronger reasoning abilities are better at interpreting complex performance feedback and designing effective sampling strategies.
This conclusion is supported by the data in Table~\ref{tab:different_model}. \texttt{Deepseek-R1} demonstrates a notable advantage, which achieves the lowest RMSE in 8 out of the 14 evaluation scenarios. For instance, when using the XGBoost model with 30 samples, \texttt{Deepseek-R1}'s RMSE of 1.660 is clearly superior to that of \texttt{GPT-4o} (2.200). Therefore, equipping the \textbf{Designer} and \textbf{Analyzer} with an LLM that has stronger reasoning abilities is a critical factor for achieving high performance.

\textbf{The choice of Generator LLM shows a non-monotonic size-performance relationship.} For the \texttt{Generator}, we investigated the relationship between model size, architecture, and effectiveness. The results, presented in Table~\ref{tab:different_gen_model}, reveal a critical trend: \textbf{there is no simple or stable relationship between a model's characteristics and its performance} in this role. \sout{This non-monotonic relationship is most evident within the \textbf{Qwen2.5} model family. For example, when training an XGBoost model with 20 samples, the smallest \texttt{Qwen2.5-7B} model achieves an RMSE of 2.177. This result is notably better than those from its much larger counterparts, \texttt{Qwen2.5-32B} (2.295) and \texttt{Qwen2.5-72B} (2.392).}
\rv{The best-performing model is inconsistent across different sample sizes and modeling techniques. For instance, the large MoE model \texttt{Mixtral-8x22B} achieves the best RMSE at 20 samples (1.818), while the medium-sized dense model \texttt{Qwen2.5-32B} performs best at 30 samples (1.743), and the small MoE model \texttt{Mixtral-8x7B} is superior at 70 samples (1.697).}
\sout{This pattern, where a smaller model can be more effective, repeats across multiple scenarios in the table.} \rv{This lack of a clear, consistently superior model} is critical for the practical application of \tool. It supports a cost-effective hybrid model strategy, as there is no evidence that a larger, more computationally expensive LLM provides a consistent advantage for the \textbf{Generator} task. Therefore, smaller, efficient LLMs can be confidently used.

\rv{\noindent \textbf{\textit{Discussion on Computational Overhead.}} These findings have implications for the cost-effectiveness and computational overhead of our approach. Our hybrid architecture strategically allocates resources based on task complexity. The \textbf{Strategy Designer} and \textbf{Performance Analyzer} roles, which demand powerful, high-reasoning models, are invoked only once per sampling iteration to analyze performance trends and formulate the next strategy. In contrast, the \textbf{Configuration Generator} is called multiple times within that same iteration to produce a stable and diverse batch of candidate configurations. As our results in Table~\ref{tab:different_gen_model} demonstrate, this high-frequency task does not benefit from larger, more complex models; smaller, efficient models are often equally or more effective. This hybrid strategy, therefore, presents a highly cost-effective solution: it concentrates computational expense on the infrequent, high-reasoning tasks while using lightweight models for the high-frequency, lower-reasoning generation task, thus minimizing overall overhead without compromising performance.}

\greybox{\textbf{RQ3 Summary:} For \tool's \textbf{Analyzer} and \textbf{Designer}, the overall performance is tied to the LLM's reasoning capability, with a powerful model securing the best results in the majority of scenarios. Conversely, the \textbf{Generator} exhibits a non-monotonic size-performance relationship.
, where a small \texttt{Qwen2.5-7B} model can achieve an RMSE over 9\% lower than its largest 72B counterpart.
}

\subsection*{RQ4: How do hyperparameter impact \tool?}
\label{sec:rq4}

\begin{figure}[tbp]
    \centering
    \includegraphics[width=0.9\linewidth]{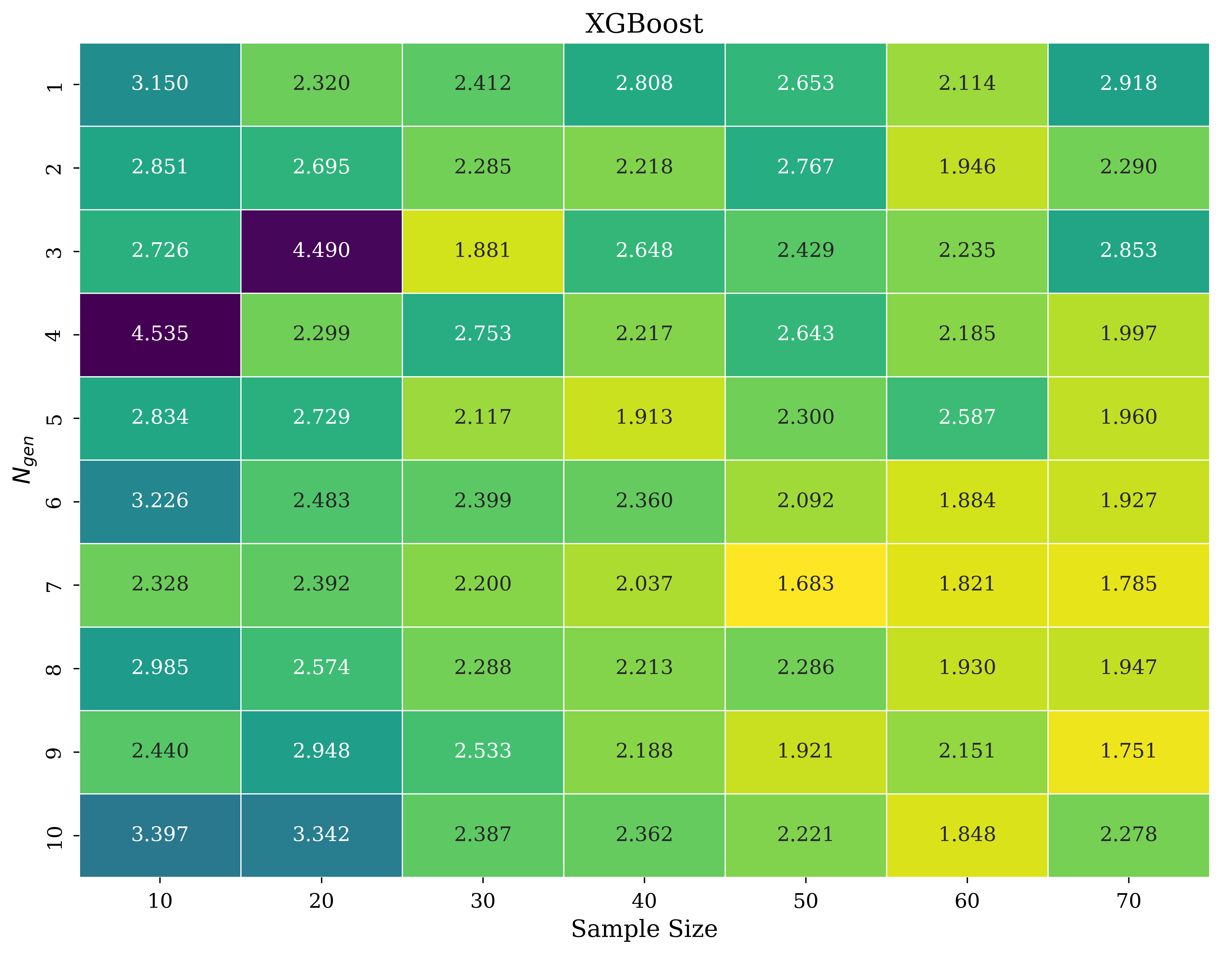}
    \caption{(RQ4) Prediction error (RMSE) of the XGBoost across different sample sizes and numbers of candidate configurations per iteration ($N_{\text{candidates}}$). Lighter shading indicates lower error, reflecting better predictive performance.}
    \label{fig:n_gen_impact}
\end{figure}
\noindent \textbf{\textit{Motivation.}}
In this section, we investigate the impact of two key hyperparameters on \tool. The first is the number of candidate configurations generated per iteration ($N_{candidates}$), which presents a critical trade-off between feedback frequency and computational cost. The second is the number of parallel generators ($N_{generators}$), which influences the diversity of generated candidates and the stability of the sampling process.

\noindent \textbf{\textit{Approach.}}
For both hyperparameter analyses, we use a consistent setup: the high-reasoning Designer and Analyzer are performed by \texttt{GPT-4o}, while the Generator uses \texttt{Qwen2.5-72B}. Each experimental setting is repeated 10 times to ensure statistical reliability. In the first experiment, we varied $N_{candidates}$ from 1 to 10 while keeping $N_{generators}$ fixed. In the second experiment, we varied $N_{generators}$ from 1 to 10 while keeping $N_{candidates}$ fixed.

\begin{figure}[tbp]
    \centering
    \includegraphics[width=0.85\linewidth]{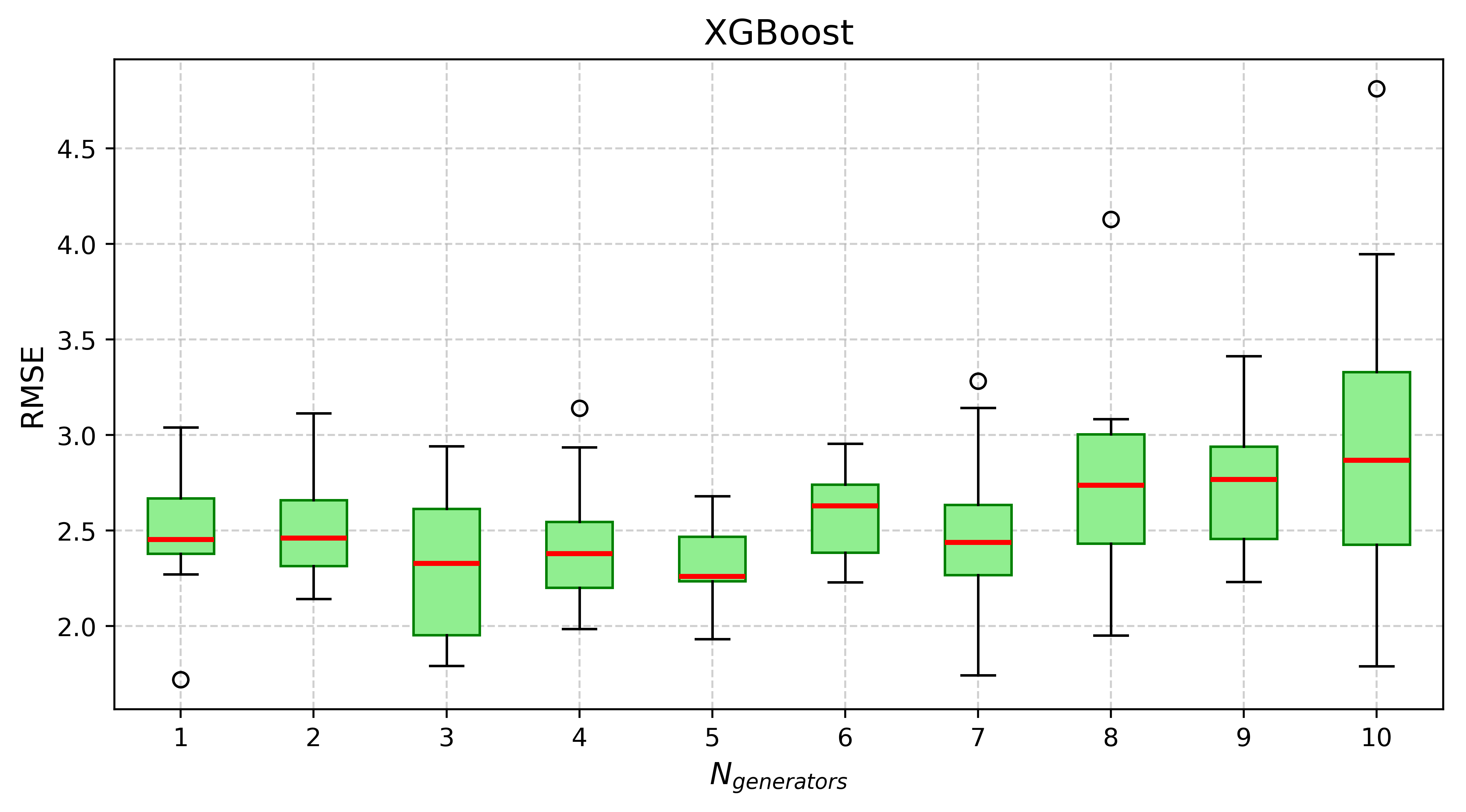}
    \caption{(RQ4) Influence of the number of parallel Configuration Generators ($N_{\text{generators}}$) on the final RMSE. Box plots summarize 10 independent runs for XGBoost.}
    \label{fig:different_gen_num}
\end{figure}

\noindent \textbf{\textit{Results.}}
\textbf{The impact of generation count ($N_{candidates}$) is complex and varies.}
The results, visualized as heatmaps in Figure~\ref{fig:n_gen_impact}, confirm that a larger candidate count does not always yield better performance, as the ideal setting for $N_{candidates}$ shifts depending on the sample size and the specific performance model. For instance, the optimal $N_{candidates}$ is 3 for a sample size of 40 but shifts to 9 for a sample size of 70. This complex relationship arises from the underlying trade-off between computational cost and feedback frequency: a small $N_{candidates}$ is expensive but provides frequent feedback, while a large count is cheaper but adapts less often. The heatmaps show the best results (brightest yellow areas) occur in a moderate range (e.g., 3 to 9), suggesting a point of diminishing returns. Therefore, a moderate value such as $N_{candidates}=7$ represents a robust and practical choice that optimally balances these competing factors.

\textbf{Using multiple generators improves sampling stability and robustness.}
As shown in Figure~\ref{fig:different_gen_num}, increasing the number of parallel \textbf{Configuration Generators} enhances the stability of sampling outcomes. When using a single generator ($N_{generators}=1$), RMSE values exhibit a wide variance, indicating unstable sampling quality. As the number of generators increases to around 3–5, both the variance and median RMSE decrease, suggesting that the ensemble-based voting mechanism effectively mitigates randomness among individual LLM outputs. However, beyond five generators, performance gains diminish and variability increases. These findings indicate that employing a moderate number of generators (3–5) provides the balance between stability, accuracy, and computational efficiency.

\greybox{
\textbf{RQ4 Summary:}
Our analysis shows moderate hyperparameter settings are optimal. We find $N_{candidates}=7$ provides the best cost-feedback balance. For $N_{generators}$, 3 to 5 provides the best trade-off for stability and performance.
}

\section{Threats to Validity}
\label{sec:threat}

\phead{Internal Validity.}
Threats to internal validity concern factors within our experiment that could influence the results. A primary factor is the hyperparameter tuning of the baseline methods. To ensure a fair comparison, we used the default parameters reported in their respective original papers where possible. Another key threat is the inherent stochasticity of the LLM used in \tool, which can lead to variability in results. We mitigate the impact of randomness across all methods by repeating each experiment multiple times and reporting the average performance. 
%During evaluation, some methods select training samples from the pruned configuration space. To ensure a fair comparison with the baselines, we assess these methods on the full configuration space instead.

\phead{External Validity.}
Threats to external validity concern the generalizability of our findings. Our evaluation is based on four open-source subject systems. Although these systems were selected to represent diverse domains (e.g., compression, databases, video encoding) following prior studies~\cite{haDeepPerfPerformancePrediction2019, xiaCoMSAModelingDrivenSampling2023, chengHINNPerfHierarchicalInteraction2023}, the results may not generalize to all types of software, particularly closed-source or unconfigurable industrial systems. 
\rv{Our approach assumes that configuration documentation is available for the target system, as it serves as the primary source of domain knowledge for search space construction. However, in scenarios where such documentation is unavailable, static or dynamic code analysis can be used to infer configuration parameters and their potential performance impacts, offering an alternative way to construct the initial search space~\cite{lillackTrackingLoadTimeConfiguration2018, li_statically_infer_2020, wangIdentifyingPerformanceSensitiveConfigurations2024}.}

\section{Conclusion}
\label{sec:conclusion}
In this paper, we conducted a comprehensive empirical study to investigate the effectiveness 
of using LLMs for the challenging task of software performance configuration sampling. To facilitate this investigation, we introduced \textbf{\tool}, a novel framework that operationalizes LLMs within an iterative feedback loop. Our study reveals that the LLM-guided approach outperforms traditional baselines in most cases, due to a dual mechanism of space pruning and feedback-driven sampling strategy refinement. Furthermore, Our study reveals that different components within \tool require LLMs with varying levels of reasoning capability and that the framework's key hyperparameter can be tuned for a practical balance between performance and cost. Overall, our work establishes LLM-guided sampling as a robust and effective new direction for performance engineering.

\bibliographystyle{ACM-Reference-Format}
\bibliography{reference}

\end{document}